\documentclass[aps,prr,twocolumn,showkeys,amsmath,amssymb,longbibliography,superscriptaddress]{revtex4-2}

\usepackage[utf8]{inputenc}
\usepackage[english]{babel}
\usepackage{graphicx}
\usepackage[table,xcdraw]{xcolor}
\usepackage{amsmath,amsfonts,amssymb,amsthm}
\usepackage[hypertexnames=false, colorlinks=true, citecolor=blue, linkcolor=red, urlcolor=teal]{hyperref}
\usepackage{braket}
\usepackage{url}
\usepackage{multirow}
\usepackage{comment}
\usepackage{enumitem}
\usepackage{booktabs}

\newtheorem{theorem}{Theorem}
\newtheorem{lemma}[theorem]{Lemma}

\usepackage[rightcaption]{sidecap}
\newcommand{\Z}{\mathbb{Z}}

\newcommand{\Q}{\mathbb{Q}}

\begin{document}
\title{
Generalized Mean-Field Ising Model as a Benchmark Family with Exact Solutions}
\author{Madita Willsch}
\thanks{These authors contributed equally to this work. \\
\href{mailto:m.willsch@fz-juelich.de}{m.willsch@fz-juelich.de}\\
\href{mailto:g.santra@fz-juelich.de}{g.santra@fz-juelich.de}}
\affiliation{J\"ulich Supercomputing Centre, Forschungszentrum J\"ulich, 52425 J\"ulich, Germany}
\affiliation{AIDAS, 52425 J\"ulich, Germany}

\author{Gopal Chandra Santra}
\thanks{These authors contributed equally to this work. \\
\href{mailto:m.willsch@fz-juelich.de}{m.willsch@fz-juelich.de}\\
\href{mailto:g.santra@fz-juelich.de}{g.santra@fz-juelich.de}}
\affiliation{J\"ulich Supercomputing Centre, Forschungszentrum J\"ulich, 52425 J\"ulich, Germany}
\altaffiliation{\protect\makebox[0pt][l]{\textsuperscript{*}}}
\author{Dennis Willsch}
\affiliation{J\"ulich Supercomputing Centre, Forschungszentrum J\"ulich, 52425 J\"ulich, Germany}
\affiliation{FH Aachen University of Applied Sciences, 52066 Aachen, Germany}
\author{Samuel Morales}
\affiliation{J\"ulich Supercomputing Centre, Forschungszentrum J\"ulich, 52425 J\"ulich, Germany}
\author{Lukas Mauth}
\affiliation{J\"ulich Supercomputing Centre, Forschungszentrum J\"ulich, 52425 J\"ulich, Germany}
\affiliation{Faculty of Mathematics and Natural Sciences, University of Cologne, 50923 K\"oln, Germany}
\author{Vrinda Mehta}
\affiliation{J\"ulich Supercomputing Centre, Forschungszentrum J\"ulich, 52425 J\"ulich, Germany}
\author{Fengping Jin}
\affiliation{J\"ulich Supercomputing Centre, Forschungszentrum J\"ulich, 52425 J\"ulich, Germany}
\author{Kristel Michielsen}
\affiliation{J\"ulich Supercomputing Centre, Forschungszentrum J\"ulich, 52425 J\"ulich, Germany}
\affiliation{AIDAS, 52425 J\"ulich, Germany}
\affiliation{Faculty of Mathematics and Natural Sciences, University of Cologne, 50923 K\"oln, Germany}
\date{\today}
\begin{abstract}

    Exactly solvable models provide opportunities to understand complex optimization problems while offering controlled benchmarks for computational methods. However, nontrivial models that combine analytical tractability with challenging optimization structure remain scarce. Here, we introduce the Generalized Mean-Field Ising (GMFI) family, a densely connected and inhomogeneous class of Ising Hamiltonians whose ground states can nevertheless be obtained exactly. This exact solution reveals a hidden low-dimensional structure and enables a complete number-theoretic classification of different system sizes and degenerate ground states. Uniquely, the GMFI family can be used to benchmark both quantum and classical optimization methods over system sizes far beyond exact diagonalization. We investigate the GMFI family using several approaches, including theoretical analysis of quantum annealing, experiments on a D-Wave quantum processor, the quantum approximate optimization algorithm (QAOA), imaginary-time evolution (ITE), greedy algorithm, simulated annealing, and a quantum-inspired greedy algorithm. Despite the GMFI model's exact solvability, it exhibits nontrivial algorithmic behavior: although the quantum-annealing gap is expected to decrease with increasing system size, it shows pronounced and unexpected system-size-dependent oscillatory signatures. These results establish GMFI models as a rare bridge between exact many-body theory, number theory, optimization, and quantum hardware, providing a scalable benchmark for identifying and understanding algorithmic limitations in classical and quantum computing.
\end{abstract}
\maketitle

\section{Introduction}
Exactly solvable models occupy a unique position in statistical~\cite{baxter1985exactly} and many-body physics~\cite{sutherland2004beautiful}. They provide rare examples in which the behavior of an interacting many-body system can be characterized analytically~\cite{baxter1985exactly,sutherland2004beautiful}, often leading to insights that extend well beyond the models themselves. Notable examples include critical phenomena in the Ising model~\cite{onsager1944,kramers41}, fractionalized spin excitations in the one-dimensional Heisenberg chain~\cite{Bethe1931,Faddeev1981}, Mott insulating behavior and spin-charge separation in the one-dimensional Hubbard model~\cite{lieb68,Essler2005,Bethe1931}, pairing correlations and superconductivity in Richardson--Gaudin models~\cite{dukelsky2004}, and topological phases and fractionalized excitations in the Kitaev honeycomb model~\cite{kitaev2006anyons}. Related analytical advances have also yielded exact results for disordered systems, notably along the Nishimori line in spin-glass models~\cite{Nishimori1980,Nishimori1981}. Beyond providing fundamental physical insights, the availability of exact analytical results also makes these models valuable benchmarks for approximate many-body theories~\cite{lipkin1965,glick1965}, classical numerical algorithms~\cite{white1992,wang2001}, and, more recently, quantum simulation and computation~\cite{salathe2015,cervera2018,miessen2024,chinnarasu2025,evered2025,vyas2026,rava2025interplay}.

In the context of quantum optimization, exactly solvable models offer a particularly valuable framework for systematically investigating the performance of optimization algorithms. Since many combinatorial optimization problems can be mapped onto finding the ground state of an Ising Hamiltonian~\cite{lucas2014}, knowledge of the exact ground state provides an unambiguous global optimum against which heuristic and approximate algorithms can be assessed~\cite{Hen2015,Wang2017,Willsch2020,Willsch2022,Mehta22}, making such models reliable benchmarks for evaluating algorithmic performance. Importantly, the analytical accessibility of the ground state does not necessarily imply that it can be efficiently reached through a given optimization protocol. Nevertheless, the physical insights of the underlying model may provide additional tools to help us understand the algorithmic performance of particular quantum optimizations~\cite{Jorg2008,Jorg_2010,altshuler2010,Seki2012,Wang2018,Bravyi2020,Wauters2020,Willsch2021JUQCSGQAOA, Mehta2021, Mehta2022,santra2024,santra2025}.

We introduce a new family of Ising optimization problems, which we refer to as the Generalized Mean-Field Ising (GMFI) problems, characterized by densely connected and spatially inhomogeneous interactions. Despite their nontrivial interaction structure, we show that their ground states can be determined exactly through an analytical construction. Furthermore, the Hamiltonian admits a compact representation in terms of two collective spin variables, revealing an underlying structure that allows us to characterize its energy landscape. Using tools from algebraic number theory, we identify special system sizes for which the continuous expression for the optimal number of up spins takes exact values, as well as those for which the ground state exhibits a two-fold degeneracy. This combination of nontrivial interaction structure and exact ground-state solvability places the GMFI model among the relatively few densely connected and inhomogeneous Ising problem families for which the global optimum is available in closed form~\cite{Mattis1976, bak1982,
Gross1984,low1994,huang2016, Rezaei2025}, making it particularly well suited for benchmarking both classical and quantum optimization algorithms.

Using this fully characterized problem family, we investigate the performance of a diverse collection of optimization algorithms, including quantum annealing~\cite{Apolloni89,Finnila1994QuantumAnnealing,KadowakiNishimori1998QuantumAnnealing,brooke99,harris2010}, the quantum approximate optimization algorithm (QAOA)~\cite{farhi2014qaoa}, and imaginary-time evolution (ITE)~\cite{Motta_2019qite,McArdle_2019varQite}, simulated annealing~\cite{kirkpatrick83}, classical~\cite{SteepestDescentGreedy} and quantum-inspired greedy search~\cite{Morales2024}. The availability of the exact ground state enables a rigorous assessment of solution quality, convergence behavior, and scalability across these different optimization paradigms. Beyond benchmarking their performance, we investigate how the underlying analytical structure of the GMFI model can help explain their behavior. In particular, for quantum annealing, we combine theoretical analysis and numerical simulations with experiments on a D-Wave quantum annealer. Similarly, we investigate the performance of QAOA, ITE, and classical and quantum-inspired optimization approaches, exploring how their behavior relates to the underlying properties of the model. These investigations reveal unexpected system-size-dependent oscillatory signatures, highlighting the value of analytical understanding in interpreting algorithmic performance.

The paper is structured as follows. In Section~\ref{sec:gmfi}, we introduce the GMFI model and discuss its defining properties. In Section~\ref{sec: direct_proof}, we prove the exact ground-state solution, while in Section~\ref{sec: mean_field_based_understanding}, we develop a mean-field-based representation to characterize the energy landscape. In Section~\ref{sec: qc_for_GMFI}, we investigate quantum optimization approaches, including a theoretical analysis of quantum annealing, experiments on a D-Wave quantum annealer, QAOA, and ITE. In Section~\ref{sec: benchmark_classical}, we examine the performance of classical and quantum-inspired optimization methods, including greedy search, simulated annealing, and a quantum-inspired greedy algorithm. Finally, in Section~\ref{sec:conclusion}, we summarize our main findings and outline possible directions for future work.

\section{Generalized Mean-Field Ising Problems}\label{sec:gmfi}

\begin{figure}
    \centering
    \includegraphics[width=\columnwidth]{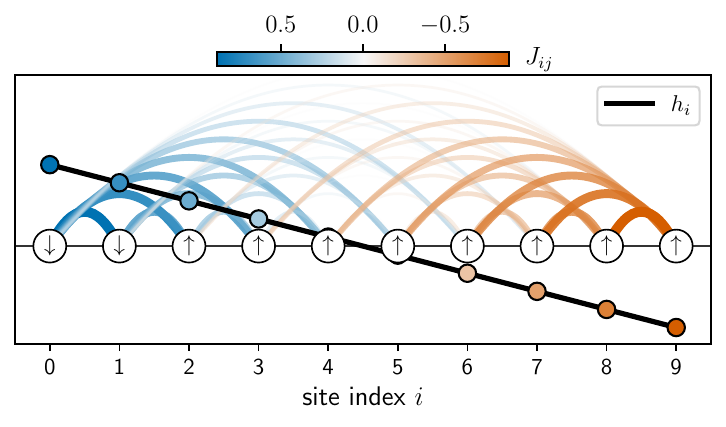}
    \caption{Schematic representation of the GMFI Hamiltonian given by Eq.~\eqref{eq:ising_gmfi} with $N=10, h=1, J=1$. The black line shows the site-dependent local interaction \(h_i\), which changes sign across the chain, with blue dots indicating positive $h_i$ and orange dots indicating negative $h_i$; the color intensity reflects its magnitude. Curved links represent long-range interactions, with blue and orange indicating positive and negative couplings $J_{ij}$, respectively. The color intensity and line thickness encode the magnitude of \(J_{ij}\), as a function of their site indices $i,j$, which favors an asymmetric ground state of the form $|\!\downarrow \cdots\uparrow \rangle $ with $\sim 79\%$ $\uparrow$ spins on the right (cf.~Eq.~\eqref{eq:Kstarlimit}), shown inside the circle on the sites.  
}
    \label{fig:hamiltonian}
\end{figure}

The Ising model~\cite{brush1967} provides a natural framework for quantum optimization, as quadratic unconstrained binary optimization (QUBO) problems can be mapped directly onto Ising Hamiltonians~\cite{AlbashLidar2018AdiabaticQuantumComputation,lucas2014}. However, finding their ground states is, in general, computationally difficult~\cite{Barahona1982IsingNPHard}.
Here, we consider a family of Ising problems in which both the local fields and the interactions possess a particular spatial structure. We refer to this family as the generalized mean-field Ising (GMFI) model, as it can be expressed in terms of mean-field-like terms (see Section~\ref{sec: mean_field_based_understanding}). The GMFI problems are defined by the Hamiltonian 
\begin{align}\label{eq:ising_gmfi}
    H =\sum\limits_{i=0}^{N-1} h_i\sigma_i^z &+ \sum\limits_{i<j} J_{ij}\sigma_i^z\sigma_j^z,\nonumber \\
    h_i = h\left(1-\frac{2i}{N-1}\right), \quad
    &J_{ij} = J\left( 1 - \frac{i+j}{N-1}\right) 
\end{align}
    where $\sigma_i^z=\ket\uparrow\!\bra\uparrow_i-\ket\downarrow\!\bra\downarrow_i$, $N>1$, and $h,J>0$.
A schematic representation of the GMFI Hamiltonian is shown in Fig.~\ref{fig:hamiltonian}.

Unlike conventional mean-field Ising models with uniform all-to-all interactions~\cite{Bapst2012QuantumMeanFieldAnnealingCurieWeiss}, the GMFI model is fully connected but strongly inhomogeneous: both the local fields and the couplings vary systematically with the positions of the spins (see Fig.~\ref{fig:hamiltonian}). In particular,
\begin{equation}
h_i\in[-h,h], \qquad J_{ij}\in \left[ -J\frac{N-2}{N-1}, J\frac{N-2}{N-1} \right].
\end{equation}
As evident, there are both positive and negative local fields and interactions, making the possibility of a fully ferromagnetic ground state unlikely. 

At first sight, it is not evident that such a densely connected and spatially inhomogeneous problem should possess a simple ground-state structure. In particular, there is no obvious reason to expect either a unique ground-state configuration or an exact analytical characterization of the low-energy spectrum. However, Mehta \textit{et al.}~\cite {Mehta22} observed an unexpectedly regular structure in the ground states found by an exhaustive search for problems with $N \leq 58$ and also in the lowest-energy samples obtained from the D-Wave Advantage\_5.1 system for problems with $N \leq 170$. These observations naturally raise the question of whether the apparent regularity is accidental or whether it originates from a hidden analytical structure of the Hamiltonian. For example, a hint of such analytical structure is provided by an exact antisymmetry under reflection of the spin indices: 

The GMFI coefficients satisfy
\begin{equation}
h_{N-1-i}=-h_i, \qquad J_{N-1-j,N-1-i}=-J_{ij}, 
\end{equation}
so under the transformation
\begin{equation}
i\mapsto N-1-i,
\end{equation}
the GMFI Hamiltonian transforms according to $H \mapsto -H$. This 
implies that every eigenstate at energy $E$ is paired with a reflected eigenstate at energy $-E$ (see Fig.~\ref{fig:energy_landscape} below). This antisymmetry structure provides a useful ingredient in the analyses presented in Sections~\ref{sec: direct_proof} and \ref{sec: mean_field_based_understanding}, helping us to uncover the eigenstate structure analytically.

\section{Exact Solution}\label{sec: direct_proof}

The ground state of the GMFI problems can be derived analytically:

\begin{theorem}
\label{th:groundstate}
The Ising problem defined in Eq.~\eqref{eq:ising_gmfi} has the ground state $|\!\downarrow \cdots \downarrow \uparrow \cdots \uparrow\rangle$ where the number $K$ of $\uparrow$ spins is given by $K=\lceil K^*\rceil$, or $K=\lfloor K^*\rfloor$, or both (``integer" or ``degenerate" case), where
    \begin{equation}
        \label{eq:IsingGroundState}
        K^* = \frac{N}{2}-\frac{{h}}{3{J}}+\frac{\sqrt{3N^2+4{h^2/J^2}}}{6}\,.
    \end{equation}
\end{theorem}

In this section, we provide the proof 
of the theorem through an elementary computation directly from the definition given by Eq.~\eqref{eq:ising_gmfi}. In the following section, we extend Theorem~\ref{th:groundstate} using ideas from a mean-field-based understanding. In particular, we further give a characterization of all $N$ for which $K=K^*\in\mathbb{N}$ (the ``integer" case, see Section~\ref{sec:ExactKStar} and App.~\ref{app:integer_k} below), and we derive a closed-form expression for all $N$ for which the GMFI ground state is degenerate (the ``degenerate" case, see Section~\ref{sec:Degeneracy} and App.~\ref{app:degeneracy_proof} below).

\textit{Proof:}
We first
show that the ground state has exactly one domain wall. To this end, consider the all-up state $\ket{\uparrow\cdots \uparrow}$ with energy $E=0$, which is easy to verify.
Let $L\in \mathbb{N}$, $1 \le L \le N$.
Flipping $L$ spins with indices $i_1, \dots, i_L$ (w.l.o.g.\ $i_1 < i_2 < \cdots < i_L$) then yields the energy change
    \begin{align}
      \Delta E = &-2\left( \sum\limits_{l=1}^L \left( h_{i_l} + \sum\limits_{j=0}^{i_l-1} J_{ji_l} + \sum\limits_{j=i_l+1}^{N-1}J_{i_lj} \right) \right. \nonumber \\
       &-\left.2\sum\limits_{l=1}^{L-1}\sum\limits_{m=l+1}^L J_{i_li_m} \right)\label{eq:energy_change_ansatz} \\
      = &\ ({J}N-2{J}L+2{h})\left[-L+\frac{2}{N-1}\sum\limits_{l=1}^L i_l \right].
      \label{eq:energy_change_direct_proof}
    \end{align}
The spin flips cause a sign change in all energy terms involving exactly one flipped spin. Counting all terms that include (at least) one flipped spin leads to double counts of terms with two flipped spins, which then have to be accounted for.

    To find the strongest negative energy change, we have to consider three cases:
    \begin{itemize}
      \setlength{\itemindent}{2em}
      \item[Case 1:] $1\le L < N/2+{h/J}$

      In this case, ${J}N-2{J}L+2{h} > 0$, and it follows that the energy is minimized (for fixed $L$) when $\sum_{l=1}^L i_l$ is minimal, i.e., the $i_l$ are taking the smallest possible values, which are $i_l = l-1,\, l=1,2,\ldots, L$, leading to a state with one domain wall.
      The minimal energy for given $L$ is then
      \begin{align*}
        \Delta E_\mathrm{min}(L) 
        &= -\frac{({J}N-2{J}L+2{h})(N-L)L}{N-1}\, .
      \end{align*}
      It is easy to see that $\Delta E_\mathrm{min}(L=1) = -{J}(N-2)-2{h} < 0$, i.e., flipping the leftmost spin (cf.~Fig.~\ref{fig:hamiltonian}) reduces the energy for all $N$. 

      \item[Case 2:] $L>N/2+{h/J}$

      In this case, ${J}N-2{J}L+2{h} < 0$, and it follows that the energy is minimized (for fixed $L$) for $\sum_{l=1}^L i_l$ maximal, i.e., the $i_l$ are taking the largest possible values, which are $i_l = N-1-L+l,\, l=1,2,\ldots, L$.
      The minimal energy for given $L$ is then
      \vspace*{-0.5\baselineskip}
      \begin{align*}
        \Delta E_\mathrm{min}(L) 
        &= \frac{({J}N-2{J}L+2{h})(N-L)L}{N-1}\, . 
      \end{align*}
      It is easy to see that $\Delta E_\mathrm{min}(L=N)=0$, i.e., flipping all bits does not yield the ground state.

      \item[Case 3:] $L=N/2+{h/J}$

      This case cannot contain the ground state as ${J}N-2{J}L+2{h} = 0$, and thus $\Delta E_\mathrm{min}(L)=0>\Delta E_\mathrm{min}(1)$.
    \end{itemize}
    We can conclude that the ground state must have exactly one domain wall and $1\le L < N$. \hfill(\!\!\!\!\!\!$\qed$)

    The next step is to show that for each state of the form $|\!\uparrow\cdots \uparrow\downarrow\cdots\downarrow\rangle$ (Case 2), there is at least one state of the form $|\!\downarrow\cdots \downarrow\uparrow\cdots \uparrow\rangle$ (Case 1) with lower energy, i.e., the ground state belongs to Case 1. 
    
    We do this by showing that for each $\tilde{L} > N/2+{h/J}$ there exists $L < N/2+{h/J}$ with $\Delta E_\mathrm{min}(L) < \Delta E_\mathrm{min}(\tilde{L})$:
    For a given such $\tilde{L}$, choose $L = N-\tilde{L}$, which fulfills $1\le L < N/2-{h/J}$. Then
    \begin{align}\label{eq: compare_two_cases_direct}
      \Delta E_\mathrm{min}(L)- \Delta E_\mathrm{min}(\tilde{L}) 
      =&-\frac{4{h}(N-\tilde{L})\tilde{L}}{N-1}< 0\, \nonumber,\\
      \Leftrightarrow \Delta E_\mathrm{min}(L) < &\,\,\Delta E_\mathrm{min}(\tilde{L})\, .
    \end{align}
    It follows that the ground state is given by a state of the form $|\!\downarrow\cdots \downarrow\uparrow\cdots \uparrow\rangle$ with $L$ spins in $\downarrow$ and $N-L$ spins in $\uparrow$ and $L< N/2+{h/J}$.\hfill(\!\!\!\!\!\!$\qed$)

    It is left to show that $L=N-K$ with $K$ defined in the Theorem.
    To this end, consider the energy difference (for $L<N/2+{h/J}$)
    \begin{align*}
      \Delta E_\mathrm{min}(L) &= -\frac{({J}N-2{J}L+2{h})(N-L)L}{N-1} \nonumber\\
      &= -\frac{({J}N^2-3{J}NL+2{J}L^2+2{h}N-2{h}L)L}{N-1}\, .
    \end{align*}
    This function has a minimum with respect to $L$ where
    \begin{align*}
      \frac{\partial \Delta E_\mathrm{min}(L)}{\partial L} = \frac{6{J}NL-\!{J}N^2-6{J}L^2-2{h}N+4{h}L}{N-1} = 0\,
    \end{align*}
    and
    \begin{align*}
    \frac{\partial ^2\Delta E_\mathrm{min}(L)}{\partial L^2} = \frac{6{J}N-\!12{J}L+4{h}}{N-1} > 0\, .
    \end{align*}
    We obtain the extrema $L_\pm = \frac{N}{2} +\frac{{h}}{3{J}} \pm \frac{\sqrt{3N^2+4{h^2/J^2}}}{6}$. From these, only $L=L_-$ satisfies $L<N/2+h/J$, and it represents a local minimum since
    $\frac{\partial ^2\Delta E_\mathrm{min}(L_-)}{\partial L^2} = \frac{2{J}\sqrt{3N^2+4{h^2/J^2}}}{N-1} > 0$.
    As $\Delta E_\mathrm{min}(L)$ is a polynomial of degree 3 with $\Delta E_\mathrm{min}(0)=\Delta E_\mathrm{min}(N)=0$, within the range $[0,N]$, this is also the global minimum.
    However, since $L_-$ is not necessarily an integer, the optimal number of $\uparrow$ spins $K$ can be $\lfloor N-L_-\rfloor =\lfloor K^* \rfloor$ or $\lceil N-L_- \rceil = \lceil K^* \rceil$ or both.\hfill \qed

\section{Mean-field-based understanding}\label{sec: mean_field_based_understanding}

The previous section establishes analytically the remarkably simple ground-state structure of the densely connected GMFI Hamiltonian. Motivated by the symmetries of the Hamiltonian, we now explore how a mean-field-based description can provide further insight into its properties. We first develop this description in Section~\ref{sec:meanfield}, and then use it to examine the energy landscape of the model in Section~\ref{Sec:EnergyLandscape}. We then use arguments from algebraic number theory to classify the ``integer" case in Section~\ref{sec:ExactKStar} and App.~\ref{app:integer_k}, and the ``degenerate" case in Section~\ref{sec:Degeneracy} and App.~\ref{app:degeneracy_proof}, respectively cf.~Theorem~\ref{th:groundstate}).

\subsection{Mean-field Representation of the GMFI Hamiltonian}
\label{sec:meanfield}
To obtain a simpler representation of the GMFI Hamiltonian, we draw intuition from the classical domain-wall structure of its ground state. In seeking collective observables that can characterize such a state, a natural starting point is the total magnetization of an arbitrary basis state $|s_0s_1\cdots s_{N-1}\rangle$, where $s_i=+1$ ($s_i=-1$) represents an $\uparrow$ ($\downarrow$) spin. The magnetization is given by
\begin{equation}
  \label{eq:M0}
  M_0=\sum_{i=0}^{N-1} s_i\,,
\end{equation}
which determines the relative populations of up and down spins. However, $M_0$ alone cannot distinguish a domain-wall configuration from other configurations with the same magnetization, as it contains no information about the spatial arrangement of the spins. This motivates the introduction of a second observable, 
\begin{equation}
  \label{eq:M1}
  M_1=\sum_{i=0}^{N-1} is_i\,,
\end{equation}
which accounts for the spin positions and attains its extremal values for domain-wall configurations at fixed $M_0$. 

As we now show, the energy $E$ of the GMFI model for an arbitrary basis state can be expressed exactly as a function of only these two observables $(M_0, M_1)$, 
providing a compact mean-field-like representation in which the ground state can be identified through energy minimization. 
Indeed, we have for the local field term in Eq.~(\ref{eq:ising_gmfi}),
\begin{equation}
     \sum_i h_is_i=h\sum_i \left( 1-\frac{2i}{N-1} \right) s_i = hM_0 - \frac{2h}{N-1} M_1\,. \nonumber
\end{equation}
Similarly, for the interaction term, we make use of the relations
\begin{align}
   M_0^2 & = 
   N+ 2 \sum_{i<j} s_i s_j\,, \nonumber \\
   M_0 M_1 
   & = \frac{(N-1)N}{2} + \sum_{i < j} (i+j) s_i s_j\,, \nonumber
\end{align}
in which we can identify all contributions to the coupling term in Eq.~(\ref{eq:ising_gmfi}).
We can thus write the energy $E$ as
\begin{align}\label{eq: mean_field_H}
    & E(M_0, M_1) = (JM_0+2h) \left(\frac{M_0}{2}-\frac{M_1}{N-1}\right)\,.
\end{align}
This reveals that the seemingly complex GMFI Hamiltonian [Eq.~\eqref{eq:ising_gmfi}] can be expressed entirely in terms of just two collective variables [Eq.~\eqref{eq: mean_field_H}]. Building on this simplified representation, we provide an alternative proof of the ground-state structure in App.~\ref{app: mean-field proof}.

\subsection{Energy Landscape}\label{Sec:EnergyLandscape}

\begin{figure*}[tbh]
    \centering
    \includegraphics[width=0.8\textwidth]{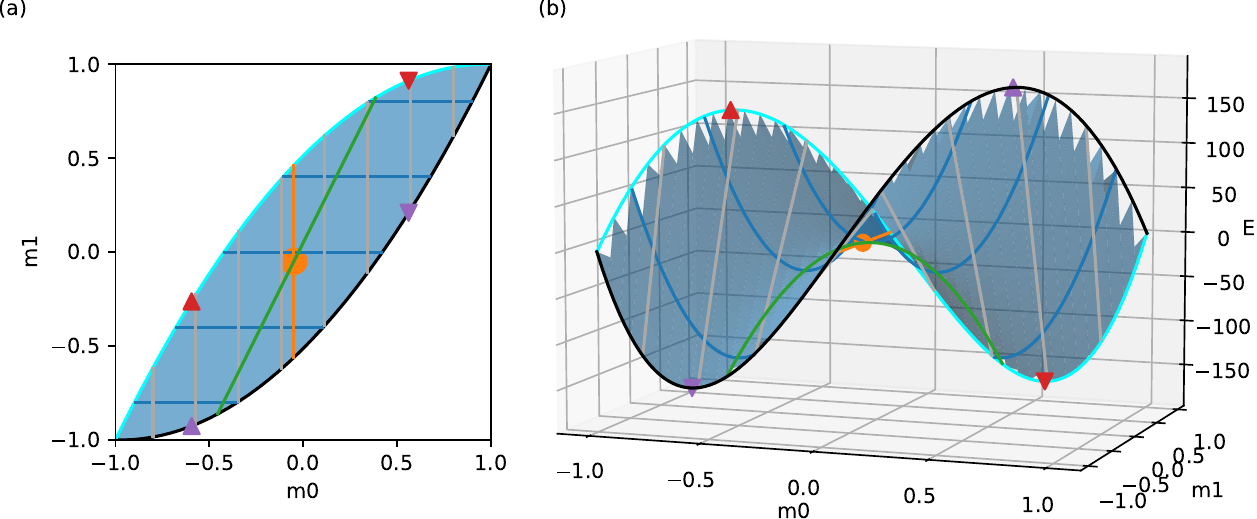}
    \caption{Characteristic energy landscape $E(m_0,m_1)$ of the GMFI problem as a function of the two normalized collective spin variables $(m_0,m_1)\in[-1,1]^2$ defined in Eqs.~(\ref{eq:m0}) and~(\ref{eq:m1}) for the case $N=40$, highlighting (a) the domain of the possible grid values $(m_0,m_1)$ and (b) the energy surface $E(m_0,m_1)$ on this domain.
    The black (cyan) line indicates the lower (upper) bound $m_1^-(m_0)$ ($m_1^+(m_0)$) given by Eq.~\eqref{eq:m1pm}.
    Gray lines (blue parabolas) indicate the functional form of $E(m_0,m_1)$ in Eq.~(\ref{eq:Em0m1}) for fixed $m_0$ ($m_1$).
    The orange line indicates the particular line in $m_1$ for which the slope as a function of $m_0$ is zero. 
    The orange circle indicates the saddle point $(m_0,m_1)=(-2/N,-2/N)$ along this line.
    The green line indicates the center of the parabolas, given by $m_0=(m_1-2/N)/2$.
    Red and purple triangles indicate the extrema of $E(m_0,m_1)$ along the lower and upper boundaries $m_1^{\mp}(m_0)$.
    Among these is the global minimum of the GMFI model (red downward-pointing triangle) given by Eq.~(\ref{eq:Em0m1min}), which corresponds to the solution $K^*$ written in Theorem~\ref{th:groundstate}.
    }
    \label{fig:energy_landscape}
\end{figure*}

In this section, we explore the two-dimensional energy landscape $E(M_0,M_1)$ of the GMFI model in terms of the collective variables $(M_0,M_1)$ (see Eq.~(\ref{eq: mean_field_H})), revealing further insights into its underlying properties.
We first note that for general spin configurations $(s_0,s_1,\ldots,s_{N-1})\in\{\pm1\}^N$, the domain of possible $(M_0,M_1)$ is given by
\begin{align}
    \label{eq:M0domain}
    M_0 &\in\{-N, -N+2, \ldots, N\}\,, \\
    \label{eq:M1domain}
    M_1 &\in\left\{-\tfrac{N(N-1)}2, -\tfrac{N(N-1)}2+2, \ldots, \tfrac{N(N-1)}2\right\}\,,
\end{align}
which can be found by analyzing the possible results of the sums in Eqs.~(\ref{eq:M0}) and~(\ref{eq:M1}).
Dividing both by their respective absolute maxima, we obtain the normalized collective variables
\begin{align}
    \label{eq:m0}
    m_0 &= \frac{M_0}{N} \in\{-1, -1+\tfrac2N, \ldots, 1\}\,, \\
    \label{eq:m1}
    m_1 &= \frac{M_1}{\frac{N(N-1)}2}\in\left\{-1, -1+\tfrac4{N(N-1)}, \ldots, 1\right\}\,,
\end{align}
on a two-dimensional grid in $[-1,1]^2$.
Note that the upper and lower extremes $(m_0,m_1)=(\pm1,\pm1)$ correspond to the unique spin configurations $\ket{\uparrow\cdots\uparrow}$ and $\ket{\downarrow\cdots\downarrow}$, respectively.

For a given value of $m_0$, which uniquely identifies the number of $\uparrow$ spins 
\begin{equation}
    \label{eq:K}
    K = \frac{N+M_0}2 = N\frac{1+m_0}2\in \{0,1,\ldots,N\}\,,
\end{equation}
all possible values of $m_1$ are constrained between two bounds $m_1^-(m_0) \le m_1 \le m_1^+(m_0)$. The lower (upper) bound corresponds to the state $\ket{\uparrow\cdots\uparrow\downarrow\cdots\downarrow}$ ($\ket{\downarrow\cdots\downarrow\uparrow\cdots\uparrow}$) with exactly $K$ $\uparrow$ spins and can be computed by evaluating Eq.~(\ref{eq:M1}) for the respective spin configurations $(s_0,s_1,\ldots,s_{N-1})=(+1,\ldots,+1,-1,\ldots,-1)$ ($(-1,\ldots,-1,+1,\ldots,+1)$). Dividing by the maximum $N(N-1)/2$, this yields the quadratic polynomials
\begin{align}
    \label{eq:m1pm}
    m_1^\mp(m_0) &= \mp\frac{N}{2(N-1)}(1-m_0^2) + m_0 \,.
\end{align}
The domain of possible $(m_0,m_1)$ respecting these bounds is shown in Fig.~\ref{fig:energy_landscape}(a).

To learn about the energy landscape of the GMFI model, we express the energy $E(M_0,M_1)$ in Eq.~\eqref{eq: mean_field_H} in terms of the normalized two-dimensional coordinates $(m_0,m_1)$. For simplicity, we consider the scaling coefficients of the GMFI model $h=J=1$. We find
\begin{align}
    \label{eq:Em0m1}
    E(m_0,m_1) = \frac{N^2}2m_0^2 + Nm_0 - Nm_1 - \frac{N^2}2m_0m_1\,.
\end{align}
This function can be interpreted in two ways:
For fixed $m_0$, $E(m_0,m_1)$ describes a line in $m_1$ whose slope is controlled by the value of $m_0$. Similarly, for fixed $m_1$, $E(m_0,m_1)$ describes a parabola in $m_0$ whose center is controlled by the value of $m_1$.
Both lines and parabolas are visualized on the surface of $E(m_0,m_1)$ in Fig.~\ref{fig:energy_landscape}(b).

Five more points of interest on the energy landscape can be computed analytically. The first is the saddle point $(m_0,m_1)=(-2/N,-2/N)$ (orange point in Fig.~\ref{fig:energy_landscape}(b)). The other four are obtained by extremizing the energy along the upper and lower bounds $m_1^{\pm}(m_0)$. To this end, note that the function $E(m_0,m_1^{\pm}(m_0))$ along these bounds is a cubic polynomial (black and cyan lines in Fig.~\ref{fig:energy_landscape}(b)), so the extremization
\begin{align}
    \frac{\mathrm{d}}{\mathrm{d} m_0} E(m_0,m_1^{\pm}(m_0)) = \frac{\partial E}{\partial m_0} + \frac{\partial E}{\partial m_1}\frac{\mathrm{d} m_1^+}{\mathrm{d} m_0} = 0
\end{align}
will yield at most two solutions along each bound (triangles in Fig.~\ref{fig:energy_landscape}(b)). Among these extrema is, in particular, also the global minimum of the energy,
\begin{align}
    \label{eq:Em0m1min}
    m_0^* = -\frac2{3N} + \sqrt{\frac4{9N^2} + \frac13},
\end{align}
corresponding to the red downward-pointing triangle Fig.~\ref{fig:energy_landscape}(b).
Note, however, that $m_0^*$ is a real number that may not be at one of the possible grid points $(m_0,m_1)$. The case in which it is, however, is classified as the ``integer" case in the following Section~\ref{sec:ExactKStar} (and App.~\ref{app:integer_k}). 

We remark that the ground state Eq.~\eqref{eq:IsingGroundState} of the GMFI model can also be obtained from this result by computing the value for the number of $\uparrow$ spins $K^*$ using Eq.~\eqref{eq:K}, yielding $K^* = (3N-2+\sqrt{3N^2+4})/6$, the same result as the one given in Theorem~\ref{th:groundstate} for $h=J=1$.
Interestingly, as $N\to\infty$, $m_0^*\to1/\sqrt3$, meaning that asymptotically, the ground state $\ket{\downarrow\cdots\downarrow\uparrow\cdots\uparrow}$ of the GMFI model has 
\begin{equation}
    \label{eq:Kstarlimit}
    K^*\to N\frac{1+1/\sqrt3}2\approx N\times79\,\%
\end{equation} $\uparrow$ spins (cf.~Fig.~\ref{fig:hamiltonian}).

Finally, a different useful representation of the energy landscape is possible by transforming $(M_0,M_1)$ to $(K,T)$, where $K$ is given by Eq.~(\ref{eq:K}) and counts the number of $\uparrow$ spins, and 
\begin{equation}
    \label{eq:T}
    T = \frac12\left(M_1+\frac{N(N-1)}2\right)\in\mathbb{N}_0
\end{equation}
represents their index sum (e.g.~a state $\ket{\downarrow\uparrow\uparrow\downarrow\uparrow}$ has $K=3$ and $T=1+2+4=7$). In these variables, the energy reads
\begin{equation}
    \label{eq:EKT}
    E(K,T) 
    = (2K-N+2)\left(K-\frac{2T}{N-1}\right)\,.
\end{equation}
This representation is particularly useful to derive a classification of degeneracies in the GMFI model in Section~\ref{sec:Degeneracy} (and in App.~\ref{app:degeneracy_proof}), as well as the determination of gaps in the performance of quantum annealing in Section~\ref{sec:theoreticalQA}.

\subsection{``Integer" Case: Closed-Form Expression for $K^*\in\mathbb N$}\label{sec:ExactKStar}

The analytical expression for the ground-state domain-wall position also reveals an interesting number-theoretic property of the GMFI model. In particular, one may ask whether there exist system sizes $N$ for which the continuous minimum $K^*$ in Eq.~\eqref{eq:IsingGroundState} is exactly an integer, thereby eliminating the need to compare the energies of its neighboring integer values. In the context of Theorem~\ref{th:groundstate}, this means that $K=\lceil K^*\rceil=\lfloor K^*\rfloor$, and in the context of the energy landscape shown in Fig.~\ref{fig:energy_landscape}(b), this means that the GMFI solution (red downward-pointing triangle) lies exactly on a grid point $(m_0,m_1)$.

We set the scaling coefficients of the GMFI model to $h=J=1$, in which case the continuous minimum is given by
\begin{equation}
    K^*=\frac{3N-2+\sqrt{3N^2+4}}{6}.
    \label{eq:Em0m1minkstar}
\end{equation}

Interestingly, the system sizes for which $K^*$ is an integer, form a discrete sequence that can be characterized using algebraic number theory. As we prove in App.~\ref{app:integer_k}, these system sizes are given by
\begin{equation}
    N_n^*=\frac{(7+4\sqrt{3})^n-(7-4\sqrt{3})^n}{\sqrt{3}}, \qquad n\in\mathbb{N}_{>0}.
    \label{eq:integer_system_sizes}
\end{equation}

The first few such system sizes are $N_n^*=8,112,1560,21728,\ldots$, with the corresponding integer solutions $K_n^*=6,88,1230,17136,\ldots$. Thus, an infinite sequence of system sizes exists for which the optimal domain-wall position is exactly an integer, without any rounding ambiguity.

\subsection{``Degenerate" Case: Degeneracies in the GMFI Model}
\label{sec:Degeneracy}

For some $N$, the ground state of the GMFI model is degenerate in the sense that two states of the form $\ket{\downarrow\cdots\downarrow\uparrow\cdots\uparrow}$ have the same minimum classical energy. In this section, we derive a closed-form expression for all such degenerate cases.

We start from the energy as a function of the number of $\uparrow$ spins $K$ and their index sum $T$ (see Eq.~\eqref{eq:EKT}). 
Since the ground state of the GMFI model lies on the upper boundary $m_1^+(m_0)$ (represented by the cyan curve in Fig.~\ref{fig:energy_landscape}(b)), we first express this boundary Eq.~\eqref{eq:m1pm} in terms of $K$ and $T$, yielding
\begin{align}
    \label{eq:TplusofK}
    T^{+}(K) = \frac{K(2N-K-1)}{2}\,.
\end{align}
Since states on this boundary are always of the form $\ket{\downarrow\cdots\downarrow\uparrow\cdots\uparrow}$ with exactly $K$ spins $\uparrow$, they have a unique representation in terms of $(K,T)$. 
Therefore, a degeneracy can only occur if two states $(K,T^+(K))$ and $(K+1,T^+(K+1))$ around the minimum (downward-pointing triangle in Fig.~\ref{fig:energy_landscape}(b)) have the same energy (note that states with $T<T^+$ for the same $K$, corresponding to grid points along the gray line from the red downward-pointing triangle to the purple upward-pointing triangle in Fig.~\ref{fig:energy_landscape}(b), will always have larger energy). 
We thus need to solve the degeneracy equation
\begin{equation}\label{eq:DegeneracyEquation}
    E(K,T^{+}(K)) = E(K+1,T^{+}(K+1)).
\end{equation}
This equation is a quadratic polynomial in $K$ and $N$, which we solve for $K$. The computation is straightforward and yields the two solutions
\begin{equation}
    \label{eq:degeneracy}
         K^\pm = \frac{3N - 5 \pm \sqrt{3N^2 + 1}}{6}.
\end{equation}
The larger of the two solutions, $K^+$, belongs to two points around the minimum (downward-pointing triangle on the cyan line in Fig.~\ref{fig:energy_landscape}
(b)). Conversely, $K^-$, belongs to the maximum (upward-pointing triangle on the cyan line). Thus, the solution for degeneracies in the ground state is given by $K^+$.

Note, however, that $K^+$ need not be an integer. Therefore, a degenerate case in the ground state of the GMFI model can only occur for those $N$ for which the solution $K^+$ in Eq.~\eqref{eq:degeneracy} is a positive integer. Following the same number-theoretic arguments as discussed in Section~\ref{sec:ExactKStar} and App.~\ref{app:integer_k}, we derive the corresponding system sizes (see App.~\ref{app:degeneracy_proof}),
\begin{equation}\label{eq:degenerate_system_sizes}
    N_n^+=\frac{(2+\sqrt{3})^{2n+1}-(2-\sqrt{3})^{2n+1}}{2\sqrt{3}}, \qquad n\in\mathbb{N}_{>0}.
\end{equation}

The first few such system sizes are $N_n^+=15,209,2911,40545,\ldots$, with the corresponding domain-wall positions $K_n^+=11,164,2295,31976,\ldots$. For each of these system sizes, the two domain-wall configurations with $K_n^+$ and $K_n^++1$ up spins have exactly the same ground-state energy.

\section{Quantum Computing for the GMFI Hamiltonian}\label{sec: qc_for_GMFI}

Having established the exact ground-state structure of the GMFI Hamiltonian and derived further properties using a mean-field representation, we now use the GMFI problems to benchmark quantum algorithms for finding the ground state.
Quantum computing has attracted considerable interest as a potential approach to solving computationally challenging problems~\cite{montanaro2016,harrow2017,abbas2024}. In this section, we explore quantum computing approaches to solving the GMFI problems, examining their theoretical properties, performance on existing quantum hardware, and prospects for implementation on emerging platforms. We begin with a theoretical analysis of quantum annealing, focusing on the behavior of the minimum energy gap - a vital factor in deciding its performance. We then benchmark quantum annealing on D-Wave hardware and investigate the performance of simulations of the quantum approximate optimization algorithm (QAOA) and imaginary-time evolution (ITE).

\subsection{Theoretical Analysis of Quantum Annealing}
\label{sec:theoreticalQA}

We first analyze ideal quantum annealing in the transverse-field Ising model~\cite{Apolloni89,Finnila1994QuantumAnnealing,KadowakiNishimori1998QuantumAnnealing} described by the time-dependent Hamiltonian
\begin{align}
    \label{eq:QAHamiltonian}
    H(s) = A(s)\sum_{i=0}^{N-1} \sigma_i^x + B(s)\left( \sum_{i<j} J_{ij} \sigma_i^z \sigma_j^z + \sum_{i=0}^{N-1} h_i \sigma_i^z \right)\,,
\end{align}
where $s\in[0,1]$ is the normalized time and the GMFI model is defined by Eq.~\eqref{eq:ising_gmfi}. Unless stated otherwise, in this section we consider a linear annealing scheme $A(s)=1-s$ and $B(s)=s$, and set the scaling coefficients in the GMFI model to $h=J=1$.

A key concept to understanding the algorithmic performance of quantum annealing is the behavior of the time-dependent \emph{spectral energy gap}
\begin{align}
    \Delta E(s) = E_1(s) - E_0(s)\,,
\end{align}
where $E_0(s)$ ($E_1(s)$) is the energy of the ground (first excited) state of the Hamiltonian $H(s)$. The central question is whether minima in the energy gap structure scale polynomially or exponentially~\cite{Bapst2012QuantumMeanFieldAnnealingCurieWeiss,AlbashLidar2018AdiabaticQuantumComputation}. As we will see below, however, for the GMFI model, a very peculiar gap behavior can be observed.

We first obtain the energy gap $\Delta E(s)$ for small $N$ below 30 by numerically diagonalizing $H(s)$ at 300 different points in time. Already at $N=29$, this requires significant computational resources and a good algorithmic approach. We use an in-house parallel implementation of the Lanczos algorithm (see~\cite{deraedt2004computational,Willsch2021JUQCSGQAOA} for more information) on the JUPITER supercomputer. The results are shown in Fig.~\ref{fig:spectral_gaps}.

An interesting structural feature of the GMFI model already emerges for $N\ge6$: There seem to be two minima between the lowest two eigenenergies (referred to as ``first'' and ``second'' gap in Fig.~\ref{fig:spectral_gaps}). Furthermore, while the first gap suggests some either polynomial or exponential scaling, the second gap shows a rather striking oscillatory behavior, suggesting that the quantum annealing performance can vary greatly between successive problem sizes $N$ (cf.~also the experimental observations in Section~\ref{sec:practicalQA} below). We systematically investigate the two gaps for larger $N$ separately.

\subsubsection{Scaling of the First Gap}
\label{sec:firstgap}

\begin{figure*}
    \centering
    \includegraphics[width=0.9\textwidth]{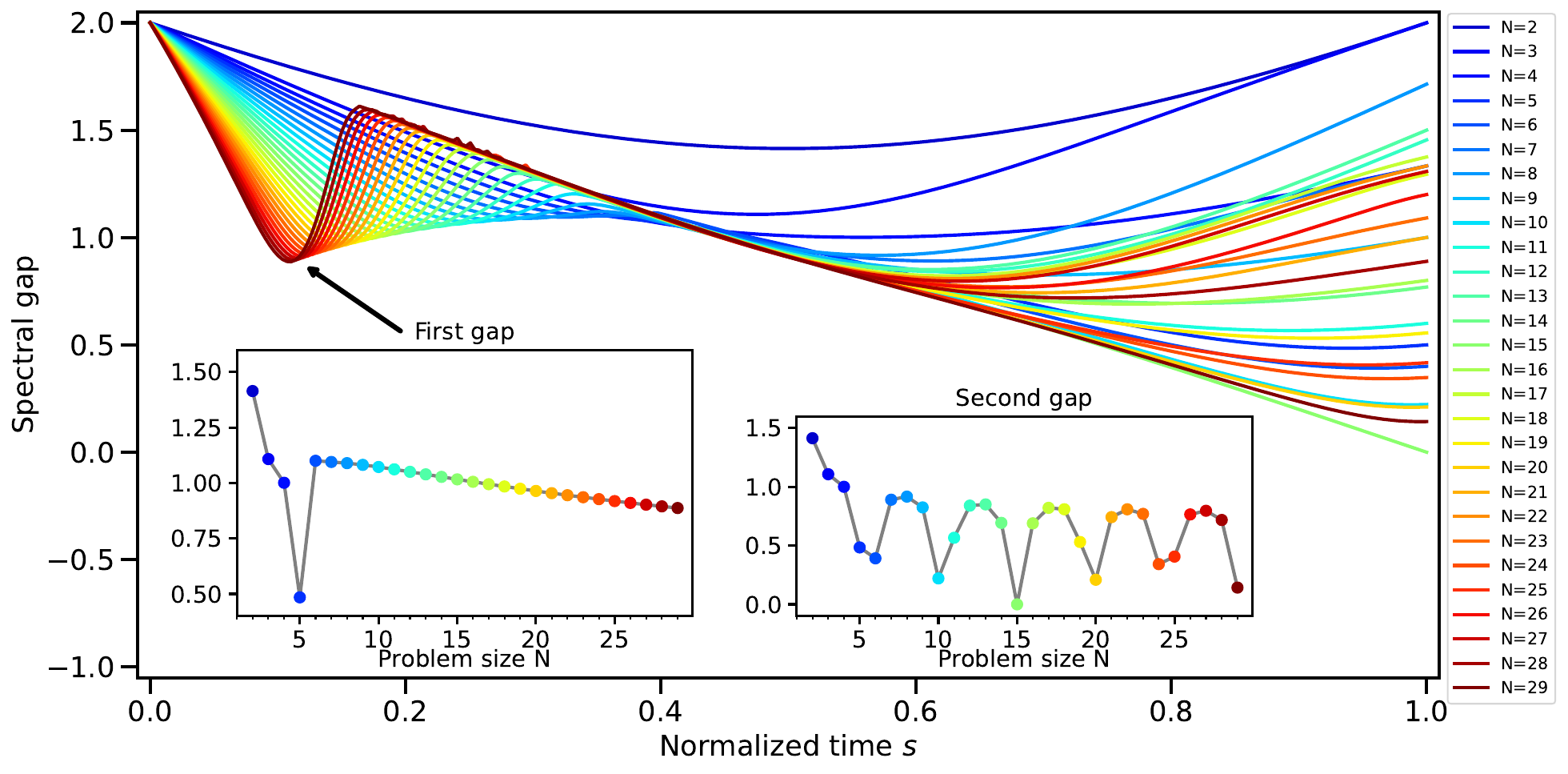}
    \caption{Spectral gap $\Delta E(s) = E_1(s)-E_0(s)$ of the quantum annealing Hamiltonian at different normalized times $s$ for problems sizes $1<N<30$.
    The data is obtained by diagonalizing the full Hamiltonian $H(s)$ given by Eq.~\eqref{eq:QAHamiltonian} for a linear annealing scheme using an in-house parallel Lanczos implementation (cf.~\cite{deraedt2004computational,Willsch2021JUQCSGQAOA}; small spikes at $s\approx0.25$ for large $N$ indicate too few Lanczos iterations that were not extended due to the high computational budget and their irrelevance for the main gaps).
    The insets show the size of the first and second gap as a function of the problem size $N$.
    For cases $N\le5$, both gaps are the same since the seperate-gap structure in the GMFI model only emerges at $N=6$.
    The degenerate case $N=15$ (cf.~Section~\ref{sec:Degeneracy}) is shown with a second gap equal to $E_1(s=1)-E_0(s=1)=0$.
    Data for the lowest three eigenenergies up to $N=15$ is shown in App.~\ref{appendixlowenergy}.
    }
    \label{fig:spectral_gaps}
\end{figure*}

\begin{figure*}
    \centering
    \includegraphics[width=\textwidth]{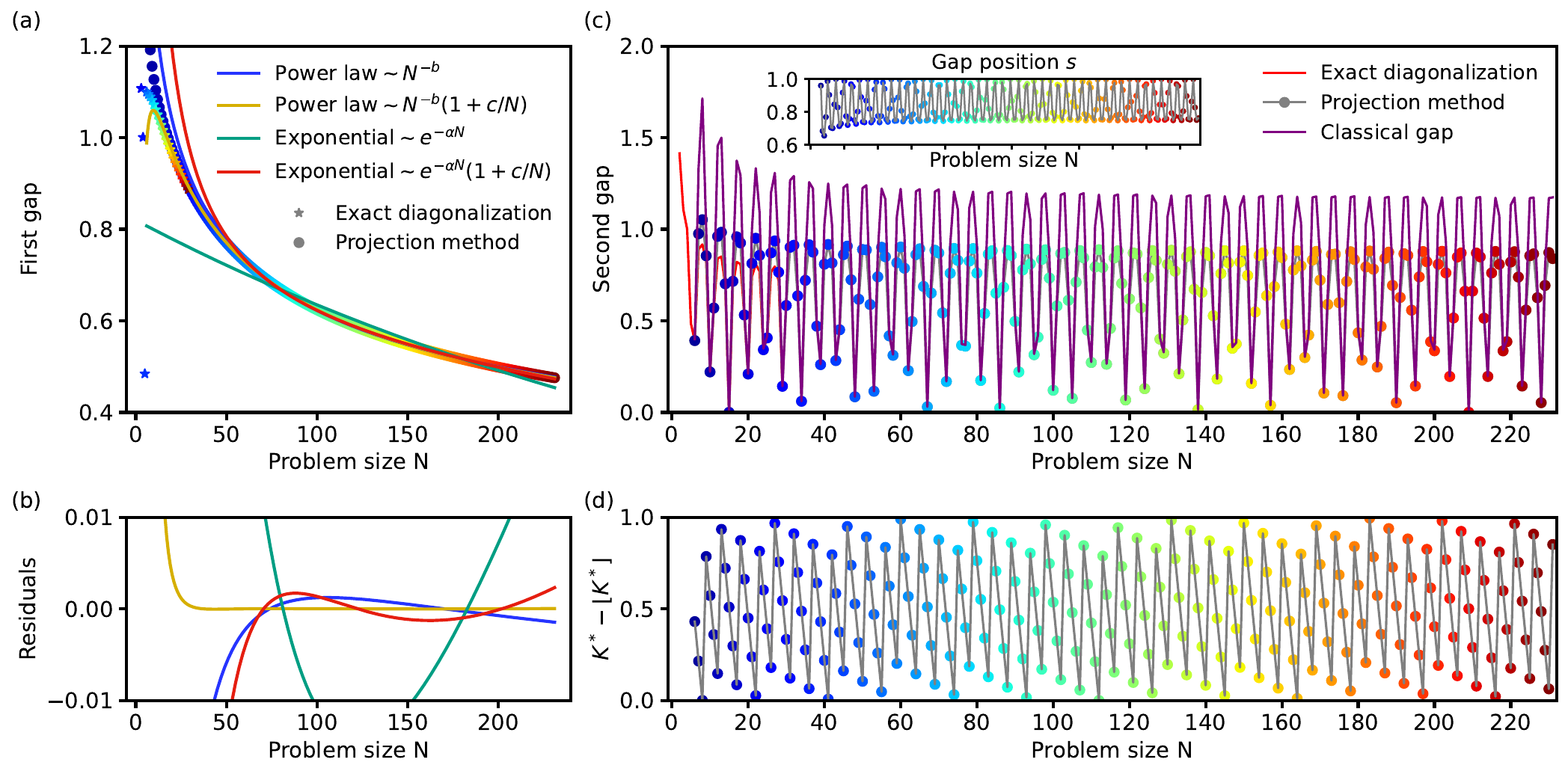}
    \caption{Characterization of the first and second gap in the quantum annealing spectrum in Fig.~\ref{fig:spectral_gaps}.
    (a) Size of the first gap obtained from exact diagonalization (stars) and the projection method described in Section~\ref{sec:firstgap} (circles) as a function of the problem size up to $N=231$ qubits.
    Lines show two polynomial and two exponential models (see legend) fitted to the gap data for all $N\ge N_0$, where $N_0$ is chosen such that the Akaike Information Criterion is minimal~\cite{Akaike1974,BurnhamAnderson2002}; specifically, $b\approx0.32$ and $N_0=55$ for the blue line, $b\approx0.34$ and $c\approx2.45$ and $N_0=34$ for the yellow line, $\alpha\approx0.0026$ and $N_0=54$ for the green line, and $\alpha\approx0.0013$ and $c\approx22.42$ and $N_0=60$ for the red line.
    (b) Residuals of the fits in (a) as a function of the problem size.
    (c) Size of the second gap obtained from exact diagonalization (red line) and the projection method (gray line). The size is anti-correlated with the position $s$, shown in the inset, of the second gap in the annealing spectrum (cf.~Fig.~\ref{fig:spectral_gaps}).
    Additionally, the purple line indicates the size of the gap in the classical case, i.e., at the end of the annealing process at $s=1$ (see App.~\ref{appendixclassicalgap} for the calculation).
    Note that the special cases $N=15$ and $N=209$ for which the GMFI ground state is degenerate (see Table~\ref{tab:degeneracies} and Theorem~\ref{thm:degeneracies}) are exactly where the circles lie on the gray line. 
    (d) Fractional part of the solution $K^*$ to the GMFI ground-state Eq.~\eqref{eq:IsingGroundState}.
    Note that the special cases $N=8$ and $N=112$ in which $K$ is an integer (see Table~\ref{tab:Kinteger} and Theorem~\ref{thm:Kinteger}) are where the circles lie exactly on the lower axis boundary at $K^*-\lfloor K^*\rfloor=0$.
    }
    \label{fig:theoretical_quantum_annealing}
\end{figure*}

The first gap in the inset of Fig.~\ref{fig:spectral_gaps} is  inconclusive w.r.t.~identifying polynomial or exponential scaling. Therefore, in this section, we devise a method to obtain the low-energy spectrum of the Hamiltonian for much larger $N$, with the goal of answering the question of polynomial vs.~exponential scaling.
We first express the quantum annealing Hamiltonian Eq.~\eqref{eq:QAHamiltonian} for a linear annealing scheme in terms of the collective spin operators $\hat M_0 = \sum_i \sigma_i^z$, $\hat M_1 = \sum_i i\sigma_i^z$, and $\hat M_x = \sum_i \sigma_i^x$, which yields (cf.~Section~\ref{sec:meanfield})
\begin{align}
    \label{eq:QAHamiltonianMx}
    H(&s) = (1-s)\hat M_x \\
    \label{eq:QAHamiltonianM0M1}
    + &s\left(  \frac{\hat M_0^2}2 + \hat M_0 - \frac{2}{N-1} \hat M_1 - \frac{1}{N-1} \hat M_0 \hat M_1 \right)\,.
\end{align}
Additionally, for given $(M_0,M_1)$, we define the state 
\begin{equation}
    \label{eq:ketM0M1}
    \ket{M_0,M_1} := \frac1{\sqrt{\Omega(N,M_0,M_1)}}\sum_{{\substack{\{s_i\}:\\\sum s_i=M_0\\\sum is_i=M_1}}} \ket{s_0s_1\cdots}\,,
\end{equation}
which is a uniform superposition over all (potentially exponentially many) states from $\ket{\downarrow\downarrow\downarrow\cdots\downarrow}$ to $\ket{\uparrow\uparrow\uparrow\cdots\uparrow}$ that have the given values of $M_0$ and $M_1$. We remark that there are $\mathcal O(N^3)$ many such states (cf.~Eqs.~\eqref{eq:M0domain} and~\eqref{eq:M1domain}).

The function $\Omega(N,M_0,M_1)$ in Eq.~\eqref{eq:ketM0M1} shall count the number of all spin configurations $(s_0,s_1,\ldots)\in\{-1,+1\}^N$ with a given $(M_0,M_1)$ (cf.~Eqs.~(\ref{eq:M0}) and~(\ref{eq:M1})). 
To compute it, we observe a useful relation to the number-theoretic integer partition function
\begin{equation}
    \label{eq:OmegaC}
    \Omega(N,M_0,M_1) = C(N,K,T)\,,
\end{equation}
which counts the number of partitions of an integer $T$ into $K$ distinct integers from $\{0,1,\ldots,N-1\}$ (note that the values $K$ and $T$ in Eq.~\eqref{eq:OmegaC} are exactly the number of $\uparrow$ spins $K$ and their index sum $T$ defined in Eqs.~\eqref{eq:K} and~\eqref{eq:T} above). 

The partition function $C(N,K,T)$ can be obtained by computing its \emph{generating} function (in the mathematical sense as a bivariate polynomial in $x$ and $y$)
\begin{equation}
    \label{eq:ZgenC}
    \mathcal Z_N(K,T) = \prod_{j=0}^{N-1} (1+xy^{j}) 
    = \sum_{KT} C(N,K,T)\, x^K y^T\,.
\end{equation}
Here, the relation between the product and the sum can be proven by observing that for each factor $(1+xy^j)$, the $1$ ($xy^j$) counts the states of the form $\ket{\updownarrow\cdots\updownarrow\downarrow\updownarrow\cdots\updownarrow}$ ($\ket{\updownarrow\cdots\updownarrow\uparrow\updownarrow\cdots\updownarrow}$) where spin $j$ is $\downarrow$ ($\uparrow$). To evaluate Eq.~\eqref{eq:ZgenC} and obtain $C(N,K,T)$, we use the so-called \texttt{flint} library for number theory~\cite{flint}.

We project the Hamiltonian $H(s)$ onto the subspace spanned by the states $\ket{M_0,M_1}$ in Eq.~\eqref{eq:ketM0M1}. 
The first part of the Hamiltonian in Eq.~\eqref{eq:QAHamiltonianMx} can be computed via
\begin{align}
    \bra{M_0',M_1'}\,\hat M_x\,&\ket{M_0,M_1} \nonumber\\
    = \sum_{j=0}^{N-1} 
\bigg( &\delta_{M_0',M_0-2} \delta_{M_1',M_1-2j} \frac{\Omega_{j\uparrow}(N,M_0,M_1)}{\sqrt{\Omega'\Omega}} \nonumber\\
+ &\delta_{M_0',M_0+2} \delta_{M_1',M_1+2j} \frac{\Omega_{j\downarrow}(N,M_0,M_1)}{\sqrt{\Omega'\Omega}} \bigg)\,,
\end{align}
where $\Omega=\Omega(N,M_0,M_1)$,  $\Omega'=\Omega(N,M_0',M_1')$, and $\Omega_{j\uparrow}(N,M_0,M_1)$ ($\Omega_{j\downarrow}(N,M_0,M_1)$) shall denote the number of spin configurations for a given $M_0,M_1$ where spin $j$ is fixed to $\uparrow$ ($\downarrow$). We compute the latter two quantities by observing that
\begin{align}
    \Omega_{j\uparrow}(N,M_0,M_1) &= C^{(j)}(N,K,T)\,,\\
    \Omega_{j\downarrow}(N,M_0,M_1) &= C(N,K,T) - C^{(j)}(N,K,T)\,,
\end{align}
where $C^{(j)}(N,K,T)$ counts the number of integer partitions similar to Eq.~\eqref{eq:OmegaC} \emph{under the constraint} that the integer $j$ is included in the partition. It can be evaluated by using that
\begin{equation}
    \frac{\mathcal Z_N(K,T)}{(1+xy^j)} = \sum_{KT} C^{(j)}(N,K,T)\, x^{K-1} y^{T-j}\,,
\end{equation}
and, as for Eq.~\eqref{eq:ZgenC}, we use the \texttt{flint} library~\cite{flint} for evaluating the coefficient.
The second part of the Hamiltonian in Eq.~\eqref{eq:QAHamiltonianM0M1} is straightforward since $\ket{M_0,M_1}$ are joint eigenstates of $\hat M_0$ and $\hat M_1$.

The projected Hamiltonian can be easily diagonalized beyond $N=200$, and we use a sparse eigensolver on the matrix in COO form~\cite{Virtanen2020SciPy} to obtain the lowest two eigenstates. Furthermore, we know that both eigenstates at $s=0$ and $s=1$ can be exactly represented in terms of the states $\ket{M_0, M_1}$, so the accuracy of the first spectral gap (whose position approaches $s=0$, cf.~Fig.~\ref{fig:spectral_gaps}) will increase with increasing $N$. In fact, a numerical computation of the first gap with exact diagonalization shows that it is already accurate up to $1\,\%$ for $N=29$ (cf.~the agreement between the stars and the circles in Fig.~\ref{fig:theoretical_quantum_annealing}(a)). 

To assess whether the gap exhibits polynomial or exponential scaling, we fit two models to the gap data as a function of the problem size $N$. The first is a power-law model of the form $aN^{-b}$ and the second is an exponential model of the form $ae^{-\alpha N}$. For both models, we also perform fits including a multiplicative finite-size correction factor $(1+c/N)$. The results are shown in Fig.~\ref{fig:theoretical_quantum_annealing}(a) and (b).

Clearly, the simple exponential model (green line) cannot describe the scaling of the first gap. However, to discriminate between the other three models, we evaluate the corresponding residuals. As Fig.~\ref{fig:theoretical_quantum_annealing}(b) shows, only the power-law model including the finite-size correction (yellow line) can describe the gap data sufficiently well. This strongly suggests that the first gap in the quantum annealing spectrum of the GMFI model exhibits polynomial scaling. We remark that this observation agrees with the polynomial closing observed in several similar quantum mean-field models such as the Curie-Weiss model (see~\cite{Bapst2012QuantumMeanFieldAnnealingCurieWeiss} for a comprehensive summary).

\subsubsection{Behavior of the Second Gap}
\label{sec:secondgap}

We first use the method from Section~\ref{sec:firstgap} to also obtain the size of the second gap up to $N=231$. The results are shown in Fig.~\ref{fig:theoretical_quantum_annealing}(c). Interestingly, the second gap shows an intriguing oscillatory behavior as a function of the problem size $N$. The size of the oscillating gap is directly anti-correlated with the position $s$ at which it occurs in the quantum annealing spectrum (see inset of Fig.~\ref{fig:theoretical_quantum_annealing}(c)). Furthermore, this oscillatory behavior also extends to the size of the ``classical gap'' at the end of the annealing spectrum (purple line).

To determine a potential period of the oscillations in the second gap, we evaluate the fractional part $K^*-\lfloor K^*\rfloor$ of the solution to the GMFI ground-state Eq.~\eqref{eq:IsingGroundState} (using $h=J=1$). As Fig.~\ref{fig:theoretical_quantum_annealing}(d) shows, the fractional part reflects the oscillatory behavior of the second gap, in the sense that it repeats every $4$--$5$ $N$ (cf.~the right inset of Fig.~\ref{fig:spectral_gaps}). This equals the number of points on the falling edge of the sawtooth. To determine this potential period analytically, we expand
\begin{align}
    \label{eq:Kstarexpand}
    K^* 
    = \frac{3+\sqrt3}6 N - \frac13 + \mathcal{O}\left(\frac1 N\right)\,.
\end{align}
Consequently, on the falling edge the fractional part of $K^*$ decreases by $\alpha = 1 - (3+\sqrt3)/6$ for every increment of $N$. The number of points on the falling edge is thus
\begin{align}
    N_{\text{points}} = \frac1\alpha = \frac{6}{3-\sqrt{3}}\approx 4.7\,.
\end{align}
Crucially, this asymptotic quasiperiod $N_{\text{points}}\not\in\mathbb Q$, so it is an irrational number and cannot constitute an exact period on the integer-valued problem sizes $N$. Moreover, the fractional parts are uniformly distributed and dense on $[0,1]$ by Weyl's equidistribution theorem (the vanishing $\mathcal{O}(N^{-1})$ in Eq.~\eqref{eq:Kstarexpand} does not alter this asymptotic property).

We thus conclude that values of $N$ for which $K^*$ approaches an integer (cf.~Section~\ref{sec:ExactKStar} and App.~\ref{app:integer_k}), and hence for which the second gap becomes arbitrarily small, occur infinitely often as $N$ increases. In other words, for any size of the second gap $\varepsilon>0$, there will be infinitely many GMFI problems at which the second gap becomes smaller than $\varepsilon$ but does not vanish. It is thus not possible to determine a scaling of this second gap as a function of $N$, or to conclude that the first gap should asymptotically be the minimum gap that would dictate the performance of quantum annealing. 

GMFI models are thus expected to exhibit an infinitely progressing \emph{oscillatory difficulty} for quantum algorithms. This unique property makes GMFI models a very interesting candidate for exploring unconventional finite-size effects and the limits of conventional scaling arguments for quantum annealing. In the following section, we investigate whether such theoretically motivated oscillations can also be observed in practical experiments.

\subsection{Quantum Annealing on D-Wave}
\label{sec:practicalQA}

\begin{figure*}[tbh]
    \centering
    \includegraphics[width=\textwidth]{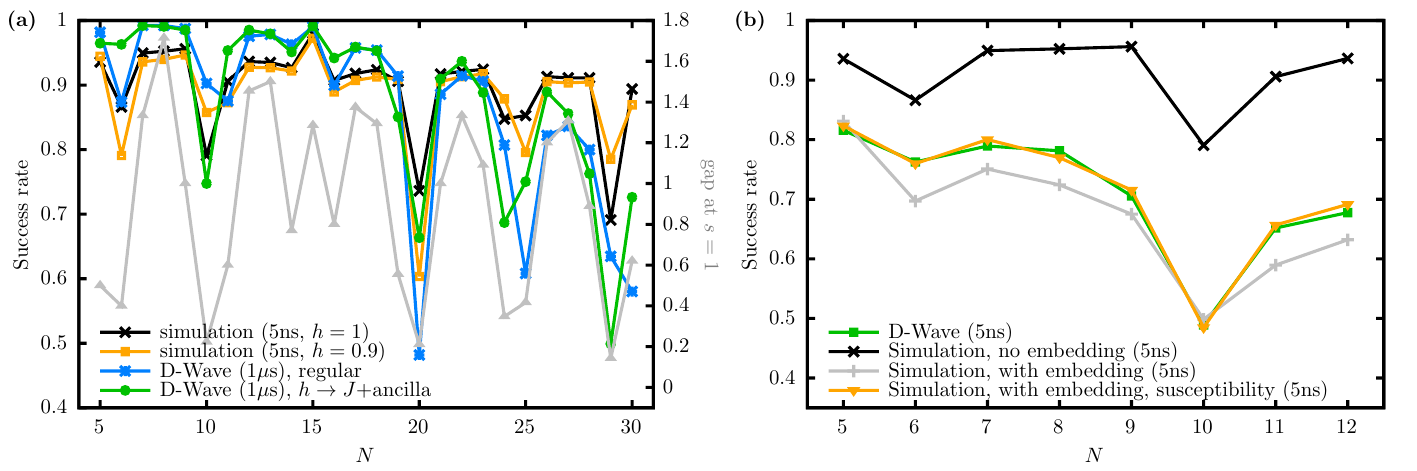}
    \caption{Quantum annealing results for the GMFI model with $h=J=1$. (a) Results from the D-Wave Advantage2\_system2.1 QPU (blue asterisks: standard annealing schedule, green circles: fast annealing schedule) in comparison to results from ideal quantum annealing simulation (black crosses). 
    For the degenerate case $N=15$, we define the success rate as the total ground-state probability.
    The simulation data follows the oscillation of the classical energy gap (gray triangles, second axis, cf.~App.~\ref{appendixclassicalgap}). While D-Wave data shows these oscillations too, there are some discrepancies (e.g., for $N=24,25$) for the data obtained with the standard annealing schedule. These discrepancies can be qualitatively reproduced when using $h=0.9$ in the simulation (orange squares).
    (b) Success rates for fast annealing with annealing time $5\,$ns for the D-Wave (green squares) and ideal simulation without taking the embedding into account (black crosses), and including the embedding step for the simulation (grey pluses). The best match is achieved when taking into account the embedding and the error induced by the qubit susceptibility (orange triangles). In this case, the data can be reproduced very nicely.
    } 
    \label{fig:success_rates_dwave}
\end{figure*}

We next investigate the GMFI model using the D-Wave Advantage2\_system2.1 quantum processing unit (QPU) located at Forschungszentrum J\"ulich. To benchmark its performance, we compare the experimentally obtained results with those from an emulation of an ideal quantum annealing process.
Our focus is on problems up to $N=30$ (since we expect the oscillatory signature with period 4.7 to surface in this range, cf.~Fig.~\ref{fig:spectral_gaps}). The advantage is that for such $N$, one can emulate the coherent quantum annealing process numerically.
This allows us to compare the data obtained from the QPU to the theoretical expectation (e.g., to search for manifestations of the oscillatory signature derived in the previous subsection) and analyze possible deviations.
This is also the major difference between our current study and the study presented in \cite{Mehta22} for the same problem family, where the focus was on the scaling of the success rate with problem size $N$, and thus problem sizes $N$ up to 170 were investigated, but for relatively large step sizes in $N$.

Since the GMFI model has an almost all-to-all connectivity, all problem instances for $N\ge 5$ require minor embedding~\cite{Choi2008} (see App.~\ref{app:embeddings} for more details).
We aim to avoid reducing the range of the $h_i$ and $J_{i,j}$ representing the problem Hamiltonian as a result of rescaling required when the largest coefficient exceeds the permitted range after fixing the chain strength. Therefore,
we set the chain strength to the minimum of the automatically deduced chain strength and 2, which is the maximum that can be set without scaling down the problem.
This leads to higher chain break ratios for larger $N$; however, instead of fixing chain breaks during post-processing, which may cause the obtained success rates to differ from those for an unembedded problem~\cite{Marshall20}, we discard these samples.
We use several different embeddings (also in parallel) as well as spin-reversal transformations to reduce the influence of the particular choice of embeddings and biases on the success rate as much as possible without the need for shimming procedures~\cite{chern23}.
Since we discard samples with chain breaks, we repeat the runs with further spin-reversal transformations until we have approximately 5000-6000 valid samples.

The results for the GMFI model with scaling coefficients $h=J=1$ obtained from the D-Wave QPU are shown in Fig.~\ref{fig:success_rates_dwave}(a). We show the success rate as a function of $N$ for the standard and the fast annealing schedules (both with an annealing time of $1\,\mu$s) in comparison to the results from ideal quantum annealing emulation by solving the time-dependent Schrödinger equation numerically~\cite{deraedt87,huyghebaert1990}. For the latter, we use the fast annealing schedule and an annealing time of $5\,$ns. 
We find that the success rates from ideal quantum annealing emulation and D-Wave data for the fast annealing schedule reflect the oscillations observed in the energy gaps at $s=1$ (see the gray line in Fig.~\ref{fig:success_rates_dwave}(a)).
Interestingly, the agreement despite the very different time scales (i.e., physical time of $1\,\mu$s and coherent emulation time of $5\,$ns) is a phenomenon that has been previously observed~\cite{Willsch2020FluxQubitsQuantumAnnealing,Willsch2020PhD,Weinberg2020,Mehta2025,Raymond2025Mar}.
The D-Wave results for the standard annealing schedule also show these oscillations, but there are some differences, e.g.,\ at $N=10,11$ and $N=24,25$, where the D-Wave data does not follow this structure.

Apart from the different annealing schedules, the D-Wave experiments also differ in how the linear terms $h_i$ are encoded.
For the standard annealing schedule (blue asterisks), the linear biases $h_i$ can be directly set.
However, for the fast annealing schedule (green circles), linear biases cannot be used~\cite{d-wave-documentation-fast-anneal}.
To still incorporate the linear terms of the problem Hamiltonian, coupling terms $J_{i,i+N}\sigma_i^z\sigma_{i+N}^z$ with additional ancilla qubits $i+N$ ($0\le i <N$) can be used~\cite{d-wave-documentation-flux-bias-offset}:
The ancilla qubits $i+N$ are strongly biased through flux biases to always attain the value $+1$.
The value of $h_i$ is then encoded in $J_{i,i+N}$ so that the terms $J_{i,i+N}\sigma_i^z\sigma_{i+N}^z$ effectively reduce to $J_{i,i+N}\sigma_i^z=h_i\sigma_i^z$.
It seems that indeed the linear bias terms are the reason for the discrepancies in the D-Wave data with the standard annealing schedule as simulation of ideal quantum annealing with $h$ set to $0.9$ but still considering the probability of the ground state of the problem with $h=1$ (orange squares) shows discrepancies at the same system sizes $N$ as the D-Wave data for the standard annealing schedule.
Thus, we assume that the strength of the linear biases is not perfectly aligned with the coupling strength, leading to a non-uniformity and effectively weaker linear biases.
This non-uniformity has previously been observed in other studies~\cite{Mehta2025}. 

\subsubsection{Quantitative comparison for fast annealing}
To assess whether we can also observe quantitative agreement between emulation and D-Wave data, we perform $5\,$ns-runs on the D-Wave QPU, using the fastest possible annealing time, because it has the highest chances of coherent evolution, and it can be emulated using reasonable compute resources.
The emulation results are shown with black crosses in Fig.~\ref{fig:success_rates_dwave}(b) and the D-Wave data with green squares.
Although the annealing time of $5\,$ns is the same in this case, we find that the D-Wave data shows a significantly lower success rate than the emulation data, deviating even more than the D-Wave data for annealing time $1\,\mu$s.

To map the almost all-to-all connected GMFI problems onto the D-Wave topology, one needs minor embedding (cf.~App.~\ref{app:embeddings}). 
Thus, it is natural to study the effect of the embedding next.
To this end, we emulate the annealing process not for the original GMFI problems, but for all the problems obtained when including the embedding step, using the same embeddings generated for the D-Wave runs but without the ancilla qubits for emulating the linear biases.
This means that for each system size $N$, we emulate $N_\mathrm{emb}$ problems incorporating about 600 different embeddings used on the D-Wave QPU.

For each embedding $j$, we compute the probability $p_{\mathrm{valid},j}$ to obtain a state that represents a valid logical qubit (i.e., no chain breaks) and the probability $p_{\mathrm{GS},j}$ to obtain the solution state.
Since for the D-Wave data we discard samples with chain breaks, we also use the probabilities $p_{\mathrm{valid},j}$ as weights in a weighted mean to obtain the overall probability of obtaining the ground state for a given system size $N$:
\begin{align}
    p_\mathrm{GS} = \frac{\sum\limits_{j=1}^{N_\mathrm{emb}} p_{\mathrm{GS},j}p_{\mathrm{valid},j}}{\sum\limits_{j=1}^{N_\mathrm{emb}}p_{\mathrm{valid},j} }\,.
\end{align}

The emulation results, including the embedding step, are shown with grey pluses in Fig.~\ref{fig:success_rates_dwave}(b).
These data are much closer to the D-Wave data than the emulation results without embedding, highlighting the effect of the embedding on the annealing process.
While this observation itself is not new~\cite{GonzalezCalaza2021,Koenz21,Vert2024,Pelofske_2025}, we want to comment on two points:
1.) It is well known and accepted that the embedding process lowers the success rate in optimization~\cite{GonzalezCalaza2021,Koenz21,Vert2024,statefactoringquantumcomputers}, but can still lead to valid results if parameters such as chain strength are chosen correctly. In experiments relying on good statistics, however, the additional embedding step may change the statistics, and thus the results significantly~\cite{Marshall20,Marshall22,Maruyama26}.
Therefore, in such cases the embedding step has to be taken with sufficient care.
2.) The D-Wave data and emulation data with embedding still do not agree very well, with the D-Wave data (green squares) surprisingly being better than the ideal emulation results with embedding. 

A possible reason could be the qubit susceptibility $\chi$, which lets the flux bias from the ancilla qubits leak onto the problem qubits, adding to the effective $h_i$ encoded through the $J_{i,i+N}$ an additional $\delta h_i \propto h_i$.
This cross-talk error is also not affected by spin-reversal transformations, leading always to an enhancement of the $h_i$.
In general, the linear bias leaks onto neighboring qubits, adding to their local fields: $h_i\sigma_i^z \mapsto (h_i + \chi\sum_k J_{ik}h_k)\sigma_i^z$~\cite{d-wave-documentation-susceptibility}.
Here we also see why spin-reversal transformations cannot remove this error: If qubit $i$ is transformed according to $\sigma_i^z\mapsto -\sigma_i^z$, the additional sign is cancelled by the analogous transformation of $h_i$ and $J_{ik}$. Similarly, if qubit $k$ is transformed, the sign cancels in the product $J_{ik}h_k$.

Assuming the flux bias on the ancilla qubits can be treated as a strong $h_{i+N}=h_{fb}$, the error $\delta h_i = \chi\sum_{k\neq i} J_{ik}h_k$ is given by $\delta h_i = \chi h_{i+N}J_{i,i+N} = \chi h_{fb}h_i$. 
However, not only do linear biases leak onto neighboring qubits through the qubit susceptibility, the qubit susceptibility also induces a next-nearest neighbor coupling~\cite{d-wave-documentation-susceptibility,King2018}
\begin{equation}
 J_{ij}\sigma^z_i\sigma^z_j \mapsto \big( J_{ij} +\chi\sum\limits_{\substack{k: (i,k)\in \mathcal{C}\wedge\\ (k,j) \in \mathcal{C}}}\!\! J_{ik}J_{kj}\big) \sigma^z_i\sigma^z_j\,.
\end{equation}
where $\mathcal{C}$ denotes the connections of the Hamiltonian.
The correction to the couplings is also invariant under spin-reversal transformation which can be seen by the same argument as for the linear correction terms.

The couplings between qubits $i$ and the ancillas $j+N$ ($j\neq i$) need to be treated differently since we can set $\sigma_{j+N}^z=1$ (samples with incorrect ancillas are discarded): $J_{i,j+N}\sigma^z_i\sigma^z_{j+N} = J_{i,j+N}\sigma^z_i$.
Thus, the effective $h_i$ are given by
\begin{align}
 &h_i\sigma_i^z \nonumber\\
 &\mapsto \big( h_i + \delta h_i + \sum\limits_{j\neq i}\big( \underbrace{J_{i,j+N}}_{=0} +\chi\!\!\!\!\!\!\sum\limits_{\substack{k: (i,k)\wedge\\ (k,j+N) \in \mathcal{C}}}\!\!\!\!\!\!\!\! J_{ik}\underbrace{J_{k,j+N}}_{h_j\delta_{kj}} \big)\big)\sigma_i^z \nonumber\\
 \label{eq:effective_hi}
 &=\big( h_i + \chi h_{fb} h_i + \sum\limits_{j\neq i}\chi J_{ij}h_j\big) \sigma^z_i\,. 
\end{align}
We find that the first term always enhances $h_i$ ($\chi$ and $h_{fb}$ are both negative).
Coupling terms where neither qubit $i$ nor qubit $j$ is an ancilla qubit are effectively described by
\begin{align}
\label{eq:effective_Jij}
 &J_{ij}\sigma^z_i\sigma^z_j \mapsto \big( J_{ij} +\chi\sum\limits_{k\neq i,j} J_{ik}J_{kj}\big) \sigma^z_i\sigma^z_j \,.
\end{align}

We use Eqs.~(\ref{eq:effective_hi})~and~(\ref{eq:effective_Jij}) to properly take into account the embeddings, which may change the actual connectivity between qubits due to qubits in chains not necessarily coupling to all qubits of the original problem.
We show the simulation results including the embedding and errors due to the susceptibility ($\chi=-0.01$, $h_{fb}=-12$) in Fig.~\ref{fig:success_rates_dwave}(b) with orange triangles.
The results fit the D-Wave data almost perfectly, showing only small deviations.

Thus, the D-Wave data from $5\,$ns runs is compatible with data from coherent time evolution.
We conclude that in the fast-annealing regime, solution quality suffers mostly from the embedding step, but the error induced by qubit susceptibility actually enhances the success rate for sampling the ground state of the GMFI model.

\subsubsection{Comparison to Boltzmann distribution}
\begin{figure*}[tbh]
    \centering
    \includegraphics[width=\textwidth]{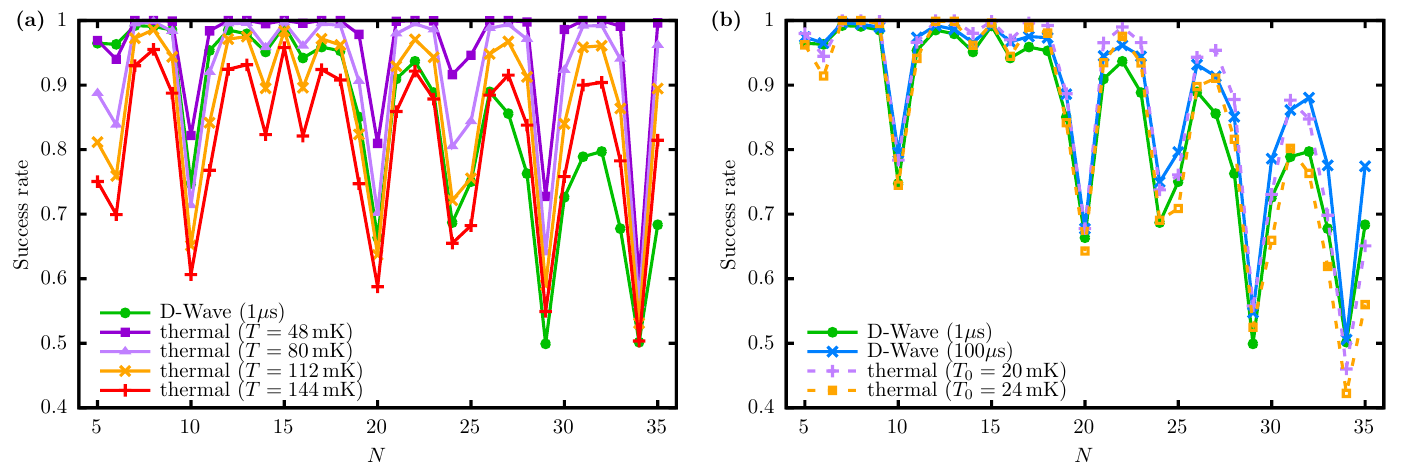}
    \caption{
    (a) The same D-Wave data as in Fig.~\ref{fig:success_rates_dwave}(a) for the fast annealing schedule in comparison to the success rate assuming the system thermalized, i.e., data is distributed according to a (classical) Boltzmann distribution for several temperatures (see legend). The success rate for thermalized data aligns as expected with the final gap size and does not show an overall decay with problem size as observed in the D-Wave data.
    (b) Success rate from D-Wave data for the fast annealing schedule using annealing times $1\,\mu$s and $100\,\mu$s in comparison to the success rate obtained from Boltzmann distributions with effective temperature depending on the average chain length which depends on $N$ for two different temperature scales.} 
    \label{fig:success_rates_dwave_comp_thermal}
\end{figure*}
For fast annealing, we could reproduce the D-Wave data by simulating the ideal coherent quantum annealing process by solving the time-dependent Schrödinger equation numerically.
For standard quantum annealing, due to thermal effects, we do not expect to be able to describe the D-Wave data by coherent time evolution from solving the time-dependent Schrödinger equation.
Thus, as a first step, we assume the system thermalized, and data sampled from this system is distributed according to the classical Boltzmann distribution (i.e. $s=1$)~\cite{Amin_2015, Mehta2025, Mehta_reverse}.
The success rate to sample the ground state from this distribution is given by $p_B=n_\mathrm{GS}\exp(-E_0\beta)/Z$, where $n_\mathrm{GS}$ denotes the degeneracy of the ground state, $E_0$ the energy of the ground state, $Z$ the partition function, and $\beta$ the inverse temperature (using $k_B=1$).

We can compute the partition function $Z=\sum_j \exp(-E_j\beta)$, where $E_j$ denotes the energy of the $j$-th eigenstate, more efficiently than brute force by making use of the energy function $\Delta E$ in Eq.~(\ref{eq:energy_change_direct_proof}) and the integer partition function $C(N,L,T)$ which gives the number of partitions of the integer $T$ into $L$ distinct integers between 0 and $N-1$.
The integer partition function $C(N,L,T)$ can be computed from the generating function
$\mathcal Z_N(L,T)$ given in Eq.~(\ref{eq:ZgenC}). 
It is straightforward to see that when expanding the product for each factor $(1+xy^j)$, the $1$ does not increase the exponents of $x$ and $y$, while the $xy^j$ adds the integer $j$ to the exponent of $y$ and $1$ to the exponent of $x$.
Thus, the exponent $L$ in $x^L$ counts how many distinct integers are added and the exponent $T$ in $y^T$ gives the sum of the added integers.
From Eq.~(\ref{eq:energy_change_direct_proof}), we find that for given $N$ and $L$, states for which $\sum\nolimits_{l=1}^L i_l= T$ is equal have the same energy $E(N,L,T) = (JN -2JL +2h)(-L+\frac{2T}{N-1})$ and $C(N,L,T)$ yields the number of these states.
Thus, instead of summing over all $2^N$ configurations $j$, we can compute the partition function as
\begin{equation}
 Z = \sum\limits_{L=0}^N \sum\limits_{T=T_\mathrm{min}}^{T_\mathrm{max}} \!\!\!\!C(N,L,T)\exp(-E(N,L,T)\beta)\,,
    \label{eq:partition_function_Z}
\end{equation}
with $T_\mathrm{min}= L(L-1)/2$ and $T_\mathrm{max}=NL-L(L+1)/2$.

We show this success rate for several temperatures in Fig.~\ref{fig:success_rates_dwave_comp_thermal}(a) in comparison to the D-Wave data obtained for annealing time $1\,\mu$s and the fast annealing schedule.
As expected, the success rates inferred from the Boltzmann distribution reflect the gap structure and thus the oscillations very well.
However, we do not observe an overall decrease in the success rate as in the D-Wave data.
We find that the D-Wave data for small $N$ rather aligns with the success rates of the ``colder" Boltzmann distribution, while for larger $N$ the agreement becomes better with the success rates from the ``warmer" Boltzmann distributions.
Assuming the system is more or less in thermal equilibrium, we would deduce that the effective temperature of the chip is higher for larger system sizes $N$.

A possible explanation could be that for larger system sizes $N$, the embeddings become more complicated and the required chains of physical qubits to achieve the required connectivity have to be longer (see also Fig.~\ref{fig:embedding_statistics} in App.~\ref{app:embeddings}).
For longer chains, freeze-out is expected to happen earlier in the annealing process~\cite{d-wave-documentation-freeze-out} so that the effective problem scale is still reduced, or in other words, the effective temperature is higher. 
Thus, we expect that the required embedding could be a reason for the decrease of the success rates with increasing $N$.

While it is easy to compute the success rate $p_B$ for the Boltzmann-distributed data because of the structure of the GMFI model (cf.\ the argument around Eq.~(\ref{eq:partition_function_Z})), it becomes intractable when including the embeddings.
Therefore, we take a different route than before and use an effective model of an $N$-dependent temperature to compute average Boltzmann distributions including the embedding step only implicitly.
To this end, we analyze the embeddings w.r.t.~the average chain length.
Assuming the value $B^*$ at the freeze-out point scales with chain length $l$ according to $B^*(l) = B(s=0.6)\cdot 0.6^{l-1}$ (where $s=0.6$ is taken to be the single-qubit freeze-out point~\cite{d-wave-documentation-freeze-out}), the effective temperature can be written as a function of $N$:
\begin{align}
  T_\mathrm{eff}(N) = T_0\frac{B(1)}{B^*(\bar l_\mathrm{chain}(N))} = T_0\frac{B(1)}{B(0.6)}0.6^{-(\bar l_\mathrm{chain}(N)-1)},
\end{align}
where $T_0$ is a constant setting the temperature scale, and $\bar l_\mathrm{chain}(N)$ is the average chain length for given $N$ obtained from averaging over all embeddings (see Fig.~\ref{fig:embedding_statistics} in App.~\ref{app:embeddings} for statistics on the generated embeddings).

We show the expected success rates for Boltzmann distributions with $N$-dependent effective temperature in Fig.~\ref{fig:success_rates_dwave_comp_thermal}(b) for two different base temperatures $T_0$, together with D-Wave data for two different annealing times.
We find that this simple model reproduces the D-Wave data qualitatively.
Although we do not claim that this model accurately describes what happens in the D-Wave device, we show that a chain-length-dependent effective temperature can reasonably reproduce the observed decrease in the success rate.

\subsection{Quantum Approximate Optimization Algorithm (QAOA)}
Given that the oscillatory difficulties of the GMFI model can be successfully observed in practical quantum annealing experiments, a natural question arises: whether the same behavior also emerges in closely related, yet distinct, quantum algorithms such as the quantum approximate optimization algorithm (QAOA)~\cite{farhi2014qaoa, zhou2020} and imaginary-time evolution (ITE)~\cite{McArdle_2019varQite,Motta_2019qite}.

Unlike quantum annealing, which follows a prescribed evolution governed by the annealing schedules $A(s)$ and $B(s)$ (see Eq.~\eqref{eq:QAHamiltonian}), QAOA replaces these schedules with $2p$ variational parameters $\boldsymbol{\beta}=(\beta_1, \beta_2, \dots, \beta_p)$ corresponding to the mixer Hamiltonian $\sum_i \sigma^x_i$ and $\boldsymbol{\gamma}=(\gamma_1, \gamma_2, \dots, \gamma_p)$ coefficients to the cost Hamiltonian ($H_C$), that are optimized through an outer-loop classical optimization procedure to minimize the variational energy $E_p(\boldsymbol{\gamma},\boldsymbol{\beta})=\langle\psi_p(\boldsymbol{\gamma},\boldsymbol{\beta})|H_C|\psi_p(\boldsymbol{\gamma},\boldsymbol{\beta})\rangle$. We employ the following strategy to find the optimal parameters: at depth \(p=1\), the energy landscape was evaluated on a uniform \(100\times100\) grid over \(\beta\in[0,\pi]\) and \(\gamma\in[0,2\pi]\) and then the lowest-energy grid point was subsequently refined using the derivative-free Constrained Optimization BY Linear Approximation (COBYLA) optimizer~\cite{Powell1994}. For each higher depth $(p>1)$, the optimized angle sequences from the preceding depth were extended by linear interpolation, providing the initial parameters for a joint optimization of all $2p$ angles. This freedom to optimize the variational parameters makes it particularly interesting to ask whether the oscillatory dependence on system size observed in quantum annealing also emerges in QAOA, even at relatively shallow circuit depths. 

To investigate whether the system-size-dependent oscillations observed in the D-Wave results persist in QAOA, we need to simulate and optimize quantum circuits across a range of system sizes. Since the computational cost of exact state-vector simulations grows exponentially with the number of qubits, we employ a Matrix Product State (MPS) implementation of QAOA~\cite{QAOATrainingPipeline} to make these calculations more tractable. During circuit simulation, we retain a maximum bond dimension of $\chi_{\mathrm{max}}=32$ and impose a relative singular-value cutoff of $10^{-12}$.

\begin{figure}
    \centering
    \includegraphics[width=\columnwidth]{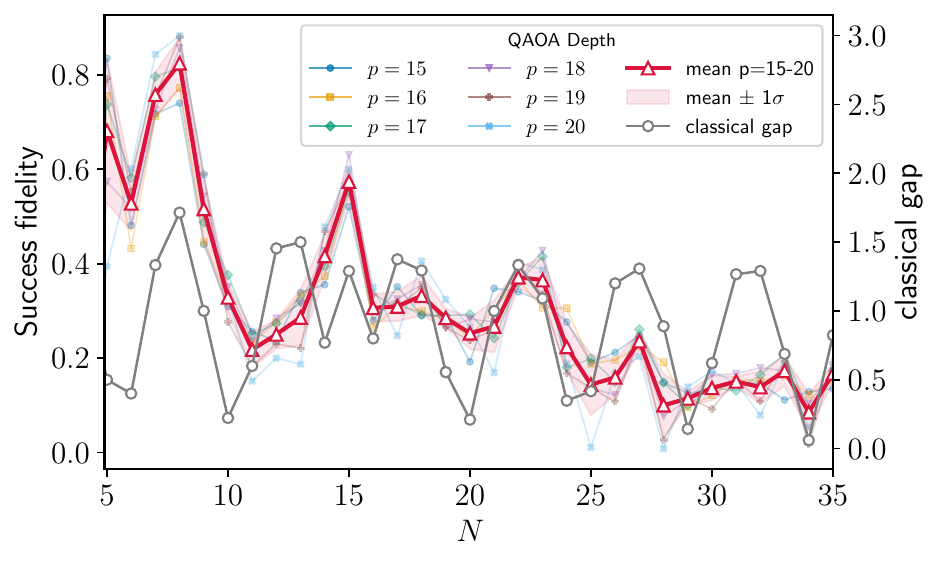}
    \caption{Success fidelity obtained using MPS-aided QAOA up to depth ($p=20$), with a maximum bond dimension $\chi_{\rm max}=32$, for system sizes from $N=5$ to $35$. The classical gap (App.~\ref{appendixclassicalgap}) is shown as a grey line on the secondary axis. Thin colored curves represent the fidelity at individual circuit depths, while the bold red curve shows the average fidelity over $p=15-20$. The QAOA fidelity reproduces the oscillatory dependence of the classical gap on the system size $N$, while its overall envelope gradually decreases with increasing $N$.
}
    \label{fig:qaoa_results}
\end{figure}

Figure~\ref{fig:qaoa_results} shows the success fidelity of the optimized QAOA states with respect to the exact ground states obtained from Theorem~\ref{th:groundstate}, for system sizes $N=5$ to $35$ and circuit depths ranging from $p=15$ to $p=20$. For the degenerate case $N=15$, we define the success fidelity as the total ground-state probability, summing the squared overlaps with both degenerate ground states, as also done for the quantum annealing results.
Across the investigated system sizes and depths, the optimized states achieve energy approximation ratios $>0.9$, defined as $r_p=E_p/E_0$, where $E_p=\langle\psi_p|H_C|\psi_p\rangle$ and $E_0<0$ is the exact ground-state energy. However, a high energy approximation ratio does not necessarily imply a high ground-state fidelity.

The success fidelity exhibits a pronounced oscillatory dependence on system size at each investigated circuit depth, broadly following the behavior of the classical energy gap. Although the fidelity varies between depths, the characteristic oscillatory pattern remains largely unchanged. To highlight this common behavior and reduce variations associated with individual circuit depths, we also consider the average fidelity over $p=15$ to $20$, shown by the red curve. 
This depth-averaged fidelity captures much of the oscillatory variation of the classical energy gap, with both quantities changing in the same direction in most of the neighboring-size comparisons over $N=5$ to $35$. Meanwhile, the overall fidelity envelope decreases with increasing system size. These results suggest that QAOA at the investigated depths captures the characteristic system-size-dependent oscillations of the GMFI model, while maintaining comparable ground-state preparation probabilities for larger systems may require greater circuit depths and improved parameter optimization.

\subsection{Imaginary-Time Evolution (ITE)}
Recently, imaginary-time evolution (ITE) has been put forward as an algorithmic avenue for ground-state preparation on quantum computers~\cite{Motta_2019qite}. Starting from a state having a finite overlap with the ground state, applying the non-unitary operator $e^{-\tau H}$ gives an exponential approach to the ground state for long times $\tau \rightarrow \infty$.

\begin{figure}
    \centering
    \includegraphics[width=\columnwidth]{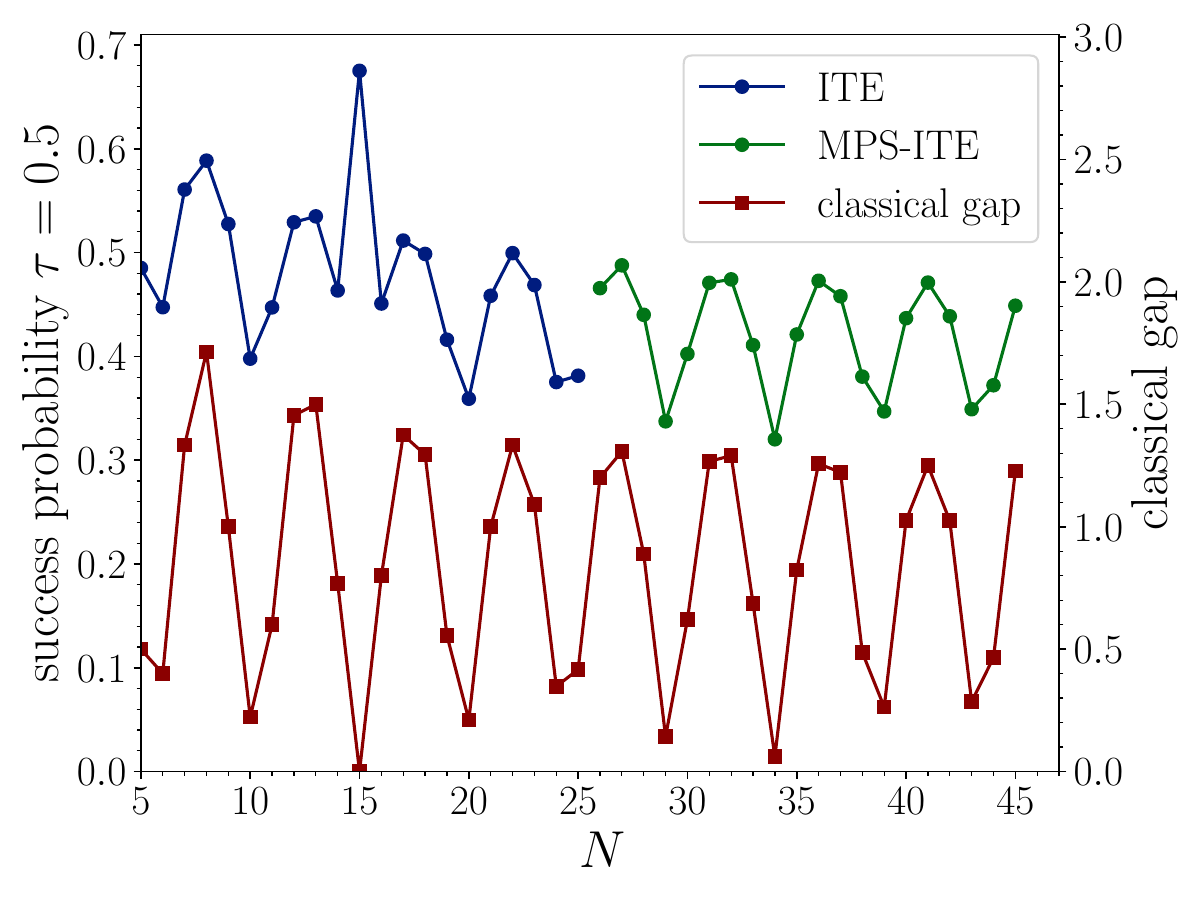}
    \caption{Success probability of the ITE state Eq.~\eqref{eq:ITE} at $\tau=0.5$ for system sizes from $N=5$ to $45$ compared to the classical gap, see App.~\ref{appendixclassicalgap}. For $N>25$, an MPS simulation with maximum bond dimension $\chi_{\rm max}=128$ is used. The oscillatory dependence of the classical gap on the system size $N$ is reflected in the behavior of ITE, except for $N=15$ corresponding to the degenerate case.
}
    \label{fig:ITE_scaling}
\end{figure}

Similar to quantum annealing and QAOA, a suitable initial state for Ising Hamiltonians is given by the product state $\ket{+\dots +}$ where $\ket{+}=(\ket{\uparrow}+\ket{\downarrow})/\sqrt{2}$, so that the ITE state at time $\tau$ is given by
\begin{equation}\label{eq:ITE}
    \ket{\psi (\tau)}= \frac{e^{-\tau H}\ket{+\dots +}}{\braket{+\dots +|e^{-2\tau H}| +\dots +}}.
\end{equation}
The non-unitary time evolution operator can be calculated analytically for the Ising Hamiltonian Eq.~\eqref{eq:ising_gmfi} as~\cite{Chai_2025iteQubo}
\begin{align}\label{eq:ITE_EvoOp}
    e^{-\tau H}=&\prod_{i=0}^{N-1} \left(\cosh(\tau h_i) - \sinh(\tau h_i) \sigma_i^z \right) \nonumber\\
    \times&\prod_{i<j} \left(\cosh(\tau J_{ij}) - \sinh(\tau J_{ij}) \sigma_i^z \sigma_j^z \right).
\end{align}
Different approaches exist to implement this kind of time evolution on quantum devices either through unitary dynamics using methods like variational circuits~\cite{McArdle_2019varQite} or double-bracket Hamiltonians~\cite{Gluza_2026dbqite}, or probabilistic implementations of nonunitary operations using ancilla-assisted measurements~\cite{Liu_2021pite, Mao_2023meas-ITE}. 

Here, we restrict ourselves to the exact statevector and MPS-aided ($\chi_{\rm max}=128$) calculation of Eq.~\eqref{eq:ITE} so that the success probability at time $\tau$ is given by
\begin{align}
    p_0(\tau)&=|\braket{\psi(\tau)|\mathrm{g.s.}}|^2=\frac{1}{1+\sum_{i\neq 0} e^{-2\tau \Delta E_i}} \nonumber\\
    &\approx 1-\sum_{i\neq 0} e^{-2\tau \Delta E_i},
\end{align}
where $\Delta E_i=E_i - E_{0}>0$ is the energy gap of the $i$-th excited state with energy $E_i$ with respect to the ground-state energy $E_0$. This corresponds to the success probability of a Boltzmann distributed state with $\beta=2\tau$.

The energy difference to the ground state is given by
\begin{align}
    &\Delta E(\tau)=E(\tau) - E_{0}
    \nonumber\\
    \approx& \sum_{i\neq \mathrm{g.s.}} \Delta E_i e^{-2\tau \Delta E_i} - \sum_{i, j\neq \mathrm{g.s.}} E_i e^{-2\tau (\Delta E_i + \Delta E_j)}.
\end{align}
Both the energy gap and the success probability are characterized by an exponential ground state convergence for long times, with the dominant contribution stemming from the classical gap $\Delta E_1$.

Figure~\ref{fig:ITE_scaling} shows the scaling of the success probability after time $\tau=0.5$. As for quantum annealing and QAOA, see also Figs.~\ref{fig:theoretical_quantum_annealing} and \ref{fig:qaoa_results}, we recover the oscillatory behaviour as expected from the classical gap.

\section{Benchmarking with classical algorithms}\label{sec: benchmark_classical}

To further investigate the GMFI problems and the signatures of its system-size dependent oscillations (cf.~Fig.~\ref{fig:theoretical_quantum_annealing}), we consider several classical and quantum-inspired algorithms. These include a greedy algorithm and simulated annealing, as well as a quantum-inspired greedy algorithm. We examine the application of these methods to the GMFI model and characterize the solutions obtained.

\subsection{Greedy Algorithm}

Solving GMFI problems with a straightforward greedy algorithm reveals important characteristics of the energy landscape shown in Fig.~\ref{fig:energy_landscape}.
For the implementation of the greedy algorithm, we use the steepest descent method~\cite{SteepestDescentGreedy}, which starts from a random classical spin configuration $(s_0,s_1,\ldots)\in\{\pm1\}^N$. It first computes the changes in energy resulting from all $N$ possible single-spin flips. In every step, it then finds and accepts the single-spin flip with the largest improvement in energy. This procedure continues until a configuration $(s_0^*,s_1^*,\ldots)\in\{\pm1\}^N$ is found for which no further single-spin flip improves the solution. This configuration (and the number of spin flips to reach it) is returned.

\begin{figure}
    \centering
\includegraphics[width=\columnwidth]{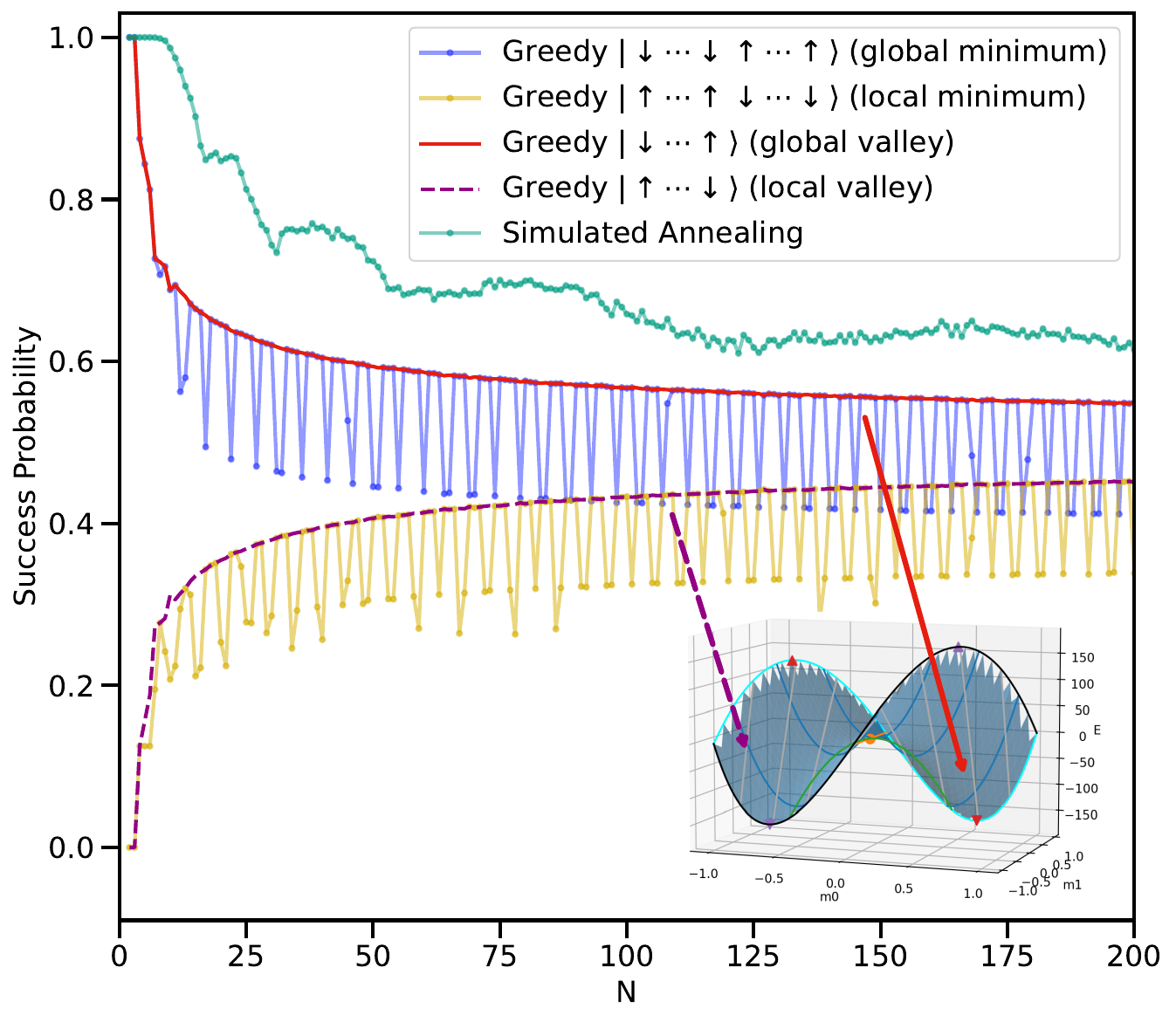}
    \caption{Success probability to find the global minimum using a classical greedy (blue) and a simulated annealing (green) algorithm as a function of the number of qubits $N$. 
    The inset shows the energy landscape from Fig.~\ref{fig:energy_landscape}.
    For the greedy algorithm, in addition to the probability of finding the global minimum (red downward-pointing triangle in the inset), the yellow line shows probability for the local minimum (purple downward-pointing triangle in the inset). 
    The solid red (dashed purple) line shows the overall probability of ending in the global (local) valley, as indicated by the two arrows on the inset.}
    \label{fig:greedy_sa}
\end{figure}

The performance of the algorithm is shown in Fig.~\ref{fig:greedy_sa}. The distinct shape of the energy landscape (cf.~Fig.~\ref{fig:energy_landscape}(b)) allows us to understand the performance of the greedy algorithm in Fig.~\ref{fig:greedy_sa}:
Given the deterministic nature of the greedy procedure, it will always deterministically converge to a local or the global minimum, and the position of this minimum is fully determined by whether there is a steepest-descent single-spin-flip path from the initial state to it. For many $N$, there are only two such minima: The global minimum of the form $|\!\downarrow \cdots \downarrow \uparrow \cdots \uparrow\rangle$ with the number of $\uparrow$ spins given by Eq.~(\ref{eq:Em0m1minkstar}), and the local minimum of the form $|\!\uparrow \cdots \uparrow \downarrow \cdots \downarrow\rangle$ (both indicated by the downward-pointing triangles in the inset of Fig.~\ref{fig:greedy_sa}).
Such $N$ are characterized by the fact that the success probabilities for the global minimum (blue dots) and the local minimum (yellow dots) add up to 1. Additionally, for these $N$, the success probabilities align with the overall probabilities of ending in the ``global valley'' (solid red line, corresponding to states characterized by $M_1>0$ which are usually of the form $|\!\downarrow \cdots \uparrow\rangle$) and those of ending in the ``local valley'' (dashed red line, corresponding to $M_1 < 0$).

However, for some $N$ there are additional local minima that cannot be improved by single-spin flips. These correspond to states of the form $|\!\downarrow \cdots \downarrow \uparrow \downarrow\uparrow\cdots \uparrow\rangle$ in the global valley (and opposite in the local valley) that are very close to the respective global and local minima, but cannot be left considering only changes from single-spin flips.

\subsection{Simulated Annealing}

For the simulated annealing algorithm, we use an in-house parallel implementation called H0SA. The software is capable of handling problems with up to four-body interactions. In the absence of three- and four-body interactions, however, like in the GMFI model, the software is equivalent to a conventional simulated annealing implementation. 

As opposed to the deterministic character of the greedy algorithm, which will always terminate in a local minimum, the simulated annealing algorithm has an additional temperature-dependent probability of leaving such local minima. As shown in Fig.~\ref{fig:greedy_sa}, it consistently achieves success probabilities above the greedy algorithm. We note that in principle, it is possible to tune the simulated-annealing parameters to always achieve a success probability close to 1, but we refrained from choosing such a setting because it would conceal the oscillatory patterns visible in the green line in Fig.~\ref{fig:greedy_sa}.


\begin{figure}
    \centering
\includegraphics[width=\columnwidth]{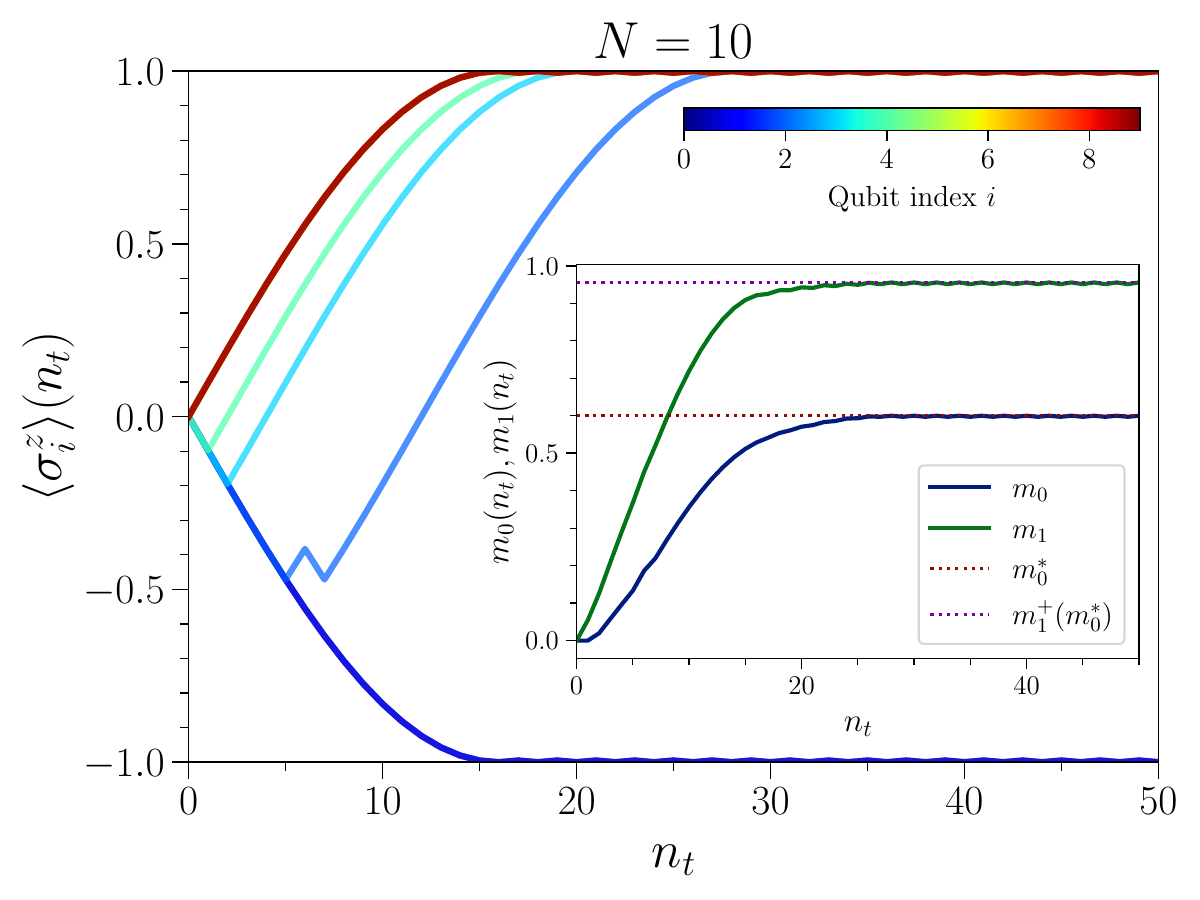}
    \caption{Evolution of the quantum greedy algorithm for $N=10$ spins. The main panel shows the evolution of the z-component of the Bloch vectors as the algorithm evolves up to $n_t=50$ time steps. The inset shows the convergence of $m_0$ and $m_1$ to the corresponding ground state values $m_0^*$ and $m_1^+(m_0^*)$ as defined in Eq.~\eqref{eq:Em0m1min} and Eq.~\eqref{eq:m1pm}.}
    \label{fig:quant_greedy}
\end{figure}

\subsection{Quantum-inspired Greedy}
Motivated from quantum protocols which gradually decrease the energy or maximize state fidelities in state preparation problems~\cite{Cerezo_2021vqa, Morales2024, Morales_2025}, we use a quantum-inspired classical algorithm to scalably prepare the ground state of the problem Hamiltonian Eq.~\eqref{eq:ising_gmfi}.

\begin{figure*}
    \centering
\includegraphics[width=\linewidth]{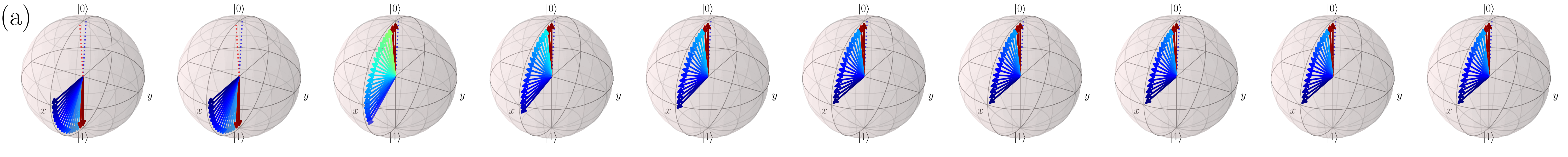}
\includegraphics[width=\linewidth]{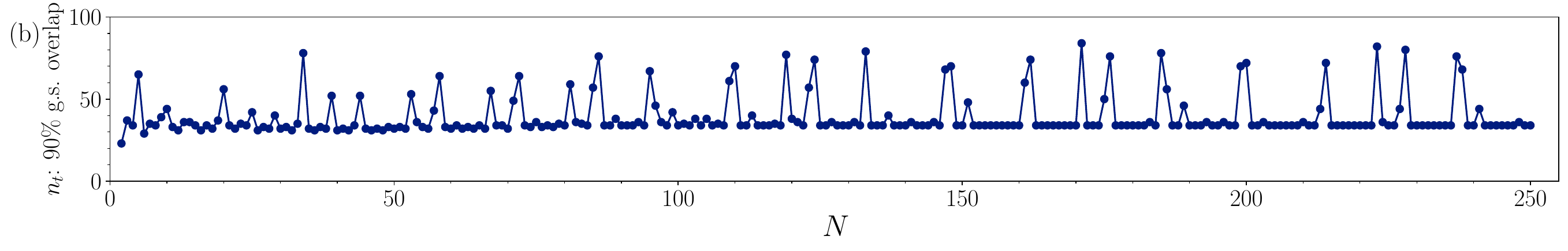}
    \caption{(a) Evolution of the quantum greedy algorithm for $N=10$ spins. The evolution of the states on the Bloch sphere is shown where each single qubit state is pure during the evolution since no entanglement is created. (b) Scaling of the time steps $n_t$ necessary to reach a ground-state probability of at least $90\%$. The rotation angle is fixed to $\phi_i=\pi/128$.}
    \label{fig:quant_greedy-bloch}
\end{figure*}

Starting from the product state $\ket{+\dots +}$ the algorithm applies single-qubit rotations $\exp({\pm i\phi_i \sigma_i^y})$ around the $y$-axis of qubit $i$ in order to reach the ground state. The direction $\eta_i=\pm 1$ of the rotation is chosen by evaluating the energy change resulting from a simultaneous rotation of two qubits $i, j$ given by
\begin{align}\label{eq:quant-greed-deltaE}
    &\braket{e^{-i \eta_i\phi_i \sigma_i^y} e^{-i \eta_{j} \phi_j \sigma_{j}^y}H e^{i \eta_i\phi_i \sigma_i^y} e^{i \eta_{j} \phi_j \sigma_{j}^y}}-\braket{H} \nonumber\\
    =&  \frac{h_i}{2} \left(\braket{\Sigma_i^{\eta_i}} \left(2+\braket{\Sigma_j^{\eta_j}} \right) - \braket{\sigma_i^z} \left(2+\braket{\sigma_j^z}\right)\right) 
    + i\leftrightarrow j,
\end{align}
with the rotated operator $\Sigma_i^{\pm}=\cos(2\phi_i)\sigma_i^z\mp \sin(2\phi_i)\sigma_i^x$. Here, we used $J_{ij}=(h_i+h_j)/2$ for $J=h$. The rotated operator satisfies $(\Sigma_i^{\pm})^2=1$, therefore in fact corresponding to an axis-rotation on the Bloch sphere.

We exploit the fact that this quantum-inspired approach keeps the product state form of initial and ground state without needing entanglement, so that $\smash{\braket{\sigma_i^\alpha\sigma_j^\beta}=\braket{\sigma_i^\alpha}\braket{\sigma_j^\beta}}$.

The single-qubit gate corresponding to the minimum value of Eq.~\eqref{eq:quant-greed-deltaE} is then applied in order to minimize the energy.
During each evolution time-step $n_t$, the whole system is split into non-overlapping nearest-neighbor pairs $(i, i+1)$ and the chosen single-qubit rotations are applied on all qubits in parallel (except for one idle qubit in the odd $N$ case).

Figure~\ref{fig:quant_greedy} shows the evolution of the algorithm for the case $N=10$. While the $z$-component of the states far away from the boundary and the collective observables $m_0$ and $m_1$ converge monotonically to their respective ground-state values, we see that spins close to the boundary might be initially turned away from their orientation in the ground state before converging to the correct value. See also Fig.~\ref{fig:quant_greedy-bloch}(a) for a visualization of the evolution on the $N$ Bloch spheres.

While for general Ising problems the algorithm will not necessarily converge to the ground state, we find that in this case the ground state is prepared successfully for any number of spins $N$ by applying a finite number $n_t$ of rotations on each spin. The angles $\phi_i$ are chosen sufficiently small to both avoid overshooting and trapping in low-energy states. Choosing $\phi_i=\pi/128$ works well up to at least $N=250$ as can be seen in Fig.~\ref{fig:quant_greedy-bloch}(b). When increasing $N$, the number of steps necessary to prepare the ground state with $90\%$ fidelity stays approximately constant around 35 steps with only a few cases needing significantly more steps. 

Further increasing $N$ can make smaller angles necessary to avoid being stuck in a low-energy state, which increases the number of necessary steps by the corresponding scaling factor compared to bigger angles. However, Fig.~\ref{fig:quant_greedy-bloch}(b) indicates that such a procedure is only needed for certain values of $N$. In particular, the occurrence of a small gap (see Fig.~\ref{fig:theoretical_quantum_annealing}(c)) does not immediately imply longer convergence times, in contrast to the quantum algorithms discussed in Section~\ref{sec: qc_for_GMFI}.

\section{Conclusion}\label{sec:conclusion}

In this work, we have introduced the Generalized Mean-Field Ising (GMFI) model, a family of densely connected and spatially inhomogeneous Ising problems whose ground states can nevertheless be determined exactly. We showed that the ground state has a simple single-domain-wall structure and that the Hamiltonian can be expressed in terms of two collective variables, revealing a hidden low-dimensional structure of the energy landscape. Moreover, we can also obtain a closed-form sequence of system sizes for integral and degenerate-ground-state cases using algebraic number theory.

The exact analytical control enables the GMFI family to serve as a benchmark for both quantum and classical optimization methods. On the quantum side, we find two distinct gap features that emerge during quantum annealing. While the first gap is consistent with polynomial closing, in agreement with behavior observed in related quantum mean-field models such as the Curie-Weiss model~\cite{Bapst2012QuantumMeanFieldAnnealingCurieWeiss}, the second gap exhibits persistent oscillations with system size. These oscillations are directly connected to the fractional part of the exact solution and therefore inherit the number-theoretic structure of the model. Since the second gap can become arbitrarily small for infinitely many system sizes $N$, no conventional scaling law in $N$ can characterize it, nor can one conclude that the first gap will asymptotically determine the performance of quantum annealing. The GMFI family thus exhibits an indefinitely recurring \emph{oscillatory difficulty}, providing a natural setting for studying unconventional finite-size effects and the limitations of standard scaling arguments in quantum optimization.

Importantly, this behavior is not restricted to an ideal theoretical construction. Indeed, we find that the same oscillatory signatures can be observed in QAOA and ITE and in physical experiments performed on a D-Wave quantum annealer. In the fast-annealing regime, the experimental behavior can be reproduced quantitatively once embedding effects and hardware susceptibility are incorporated, illustrating how an exactly solvable model can help distinguish intrinsic algorithmic structure from hardware-induced effects.

Similarly, for the classical greedy, simulated annealing, and quantum-inspired greedy algorithms, we see variations in the performance as a function of system size, which can be related to the structure of the energy landscape and the emergence of additional local minima.

The GMFI model thus provides a rare setting in which exact many-body theory, number theory, optimization algorithms, and quantum-hardware experiments can be studied within the same framework. Its mean-field structure also motivates further investigation of its finite-temperature properties and the possible relation between its quantum gap structure and thermal phase behavior. On the experimental side, its dense connectivity makes trapped-ion quantum processors~\cite{wineland2003,leibfried2003,haffner2008,bruzewicz2019} particularly interesting, as their native long-range interactions could reduce or avoid the substantial embedding overhead encountered on sparsely connected annealing hardware. In particular, magnetic-gradient-induced coupling (MAGIC) architectures offer a promising direction, combining all-to-all Ising interactions~\cite{mintert2001ion,khromova2012} with individual microwave addressing~\cite{johanning2009addressing}, low crosstalk~\cite{piltz2014crosstalk}, and high-fidelity microwave-based control~\cite{harty2014}. These capabilities, together with the simple structure of the GMFI Hamiltonian, motivate exploring its possible implementation on such platforms.

\begin{acknowledgments}

M.W., G.C.S., D.W., S.M., L.M., V.M., and F.J. acknowledge support from the project Entwicklungspartnerschaft Ionenfallen-Quantencomputer in NRW (EPIQ), funded by the Ministerium für Kultur und Wissenschaft des Landes Nordrhein-Westfalen (MKW NRW).
M.W. and D.W. acknowledge support from the project Jülich UNified Infrastructure for Quantum Computing (JUNIQ) that has received funding from the German Federal Ministry of Education and Research (BMBF) and the Ministry of Culture and Science of the State of North Rhine-Westphalia.
M.W., D.W., and L.M. acknowledge support from the project QEC4QEA that has received funding from the European High Performance Computing Joint Undertaking (EuroHPC JU) under Grant Agreement No 101194322. D.W. acknowledges support
from the project “Quantum-based Energy Grids (QuGrids),” funded under the programme “Profilbildung 2022,” an initiative of the Ministry of Culture and Science of the State of North Rhine-Westphalia. This work was supported by Q-NEKO through the EuroHPC Joint Undertaking under Grant Agreement No. 101241875 and its members, including top-up funding by the Government of Japan, and was co-funded by the European Union.
The authors thank the Gauss Centre for Supercomputing e.V. (GCS) for providing computing time on the Supercomputer JUPITER at the Jülich Supercomputing Centre. JUPITER is supported by the EuroHPC JU and GCS through funding by the European Commission, the German Federal Ministry of Research, Technology and Space, and the Ministry of Culture and Science of the State of North Rhine-Westphalia.
The authors gratefully acknowledge computing time on the supercomputer JURECA~\cite{JURECA} at Forschungszentrum Jülich under grant no.~eqip.
The authors gratefully acknowledge the Jülich Supercomputing Centre (\url{https://www.fz-juelich.de/jsc}) for funding this project by providing computing time on the D-Wave Advantage2™ System JUPSI through the Jülich UNified Infrastructure for Quantum computing (JUNIQ). 

\end{acknowledgments}

\bibliography{bib}

\section{Data Availability}
A repository that demonstrates the generation of problems from the GMFI model family will become publicly accessible as part of the JUNIQ Benchmark Suite at \url{https://jugit.fz-juelich.de/juniq/benchmark-suite/qubo-ising-benchmarks/generalized-mean-field-ising}. The data that support the findings of this article are available upon reasonable request from the authors.

\appendix
\section{Mean-field based proof of Theorem.~\ref{th:groundstate}}\label{app: mean-field proof}

In this appendix, we present an alternative proof of Theorem~\ref{th:groundstate} using the mean-field representation of the GMFI Hamiltonian introduced in the Section.~\ref{sec:meanfield}. The proof proceeds in two steps: first, we show that the ground state must have a single-domain-wall configuration, with all down spins to the left and all up spins to the right; second, we determine the optimal position of this domain wall by minimizing its energy. 

Starting from Eq.~\eqref{eq: mean_field_H}, the problem reduces to identifying the values of $M_0$ and $M_1$ that minimize the energy. For a fixed value of \(M_0\), the energy depends on the spatial arrangement of the spins only through \(M_1\). From Eq.~\eqref{eq: mean_field_H}, we have
\begin{equation}
    \frac{\partial E}{\partial M_1}
    =
    -\frac{JM_0+2h}{N-1}.
\end{equation}
This gives three possible cases:
\begin{itemize}
    \item Case 1. $\frac{\partial E}{ \partial M_1} < 0 \Rightarrow M_0 >-2h/J$ (to lower energy, maximize $M_1$)
    \item Case 2:  $\frac{\partial E}{ \partial M_1} >0 \Rightarrow M_0 <-2h/J$ (to lower energy, minimize $M_1$)
    \item Case 3: $\frac{\partial E}{ \partial M_1} = 0 \Rightarrow M_0=-2h/J$ (the energy is independent of \(M_1\) at fixed \(M_0\), leading to degeneracies.)
\end{itemize}

For a fixed value of $M_0$, minimizing the energy therefore requires either maximizing or minimizing $M_1$, depending on the sign of $JM_0+2h$. Both extrema correspond to single-domain-wall configurations, in which all up spins occupy either the rightmost or the leftmost sites.

It remains to determine which of these two configurations gives the global ground state. Consider two possibilities, both with $K$ number of $\uparrow$:
\begin{align}
|\psi_1(K)\rangle&=|\downarrow_0\cdots\downarrow_{N-K-1} \uparrow_{N-K}\cdots\uparrow_{N-1}\rangle,\\
|\psi_2(K)\rangle &=|\uparrow_0\cdots\uparrow_{K-1}\downarrow_K\cdots\downarrow_{N-1}\rangle.
\end{align}
For any configuration of type $|\psi_2(K)\rangle$, we consider its globally spin-flipped counterpart $|\psi_1(N-K)\rangle$. Since the two-body interactions are invariant under a global spin flip, while the local-field contribution changes sign, their energies satisfy
\begin{align}
&\langle\psi_1(N-K)|H|\psi_1(N-K)\rangle-\langle\psi_2(K)|H|\psi_2(K)\rangle \nonumber\\
&\qquad = -\frac{4hK(N-K)}{N-1}<0,
\end{align}
for $h>0$ and $0<K<N$.

Thus, every configuration of the second type has a configuration of the first type with strictly lower energy. Consequently, the ground state must belong to the first family, with all down spins on the left and all up spins on the right.

In Case 3, where $JM_0+2h=0$, the energy becomes independent of $M_1$, allowing multiple spin configurations to have the same energy. However, it leads Eq.~\eqref{eq: mean_field_H} to $E=0$, whereas a single-domain-wall configuration with $K=N-1$ has a lower energy, i.e., $E(N,N-1)=-J(N-2)-2h<0$ for $J,h>0$ and $N\geq2$. Therefore, Case 3 cannot correspond to the global ground state and can be excluded from further consideration.

Having established the form of the ground state, the remaining task is to determine the optimal position of the domain wall, or equivalently, the number $K$ of up spins. For the configuration $|\psi_1(K)\rangle$, the collective variables are given by
\begin{align*}
    M_0 &= -(N-K)+K=2K-N; \\
    M_1 &= \frac{N(N-1)}{2}-2\frac{(N-K)(N-K-1)}{2}\,.
\end{align*}
From here, we obtain the energy as
\begin{equation}
    E(N,K)=-\frac{K(N-K)}{N-1}[J(2K-N)+2h]\,.
\end{equation}
Taking the derivative with respect to $K$ (assuming continuity), we obtain
\begin{equation}
    \frac{\partial E}{\partial K}=0 \Rightarrow J(N^2-6NK+6K^2)+2h(2K-N)=0\,.
\end{equation}
Solving for $K$, we find
\begin{equation}
    {K}^*_{\pm}=\frac{N}{2}-\frac{h}{3J}\pm\frac{\sqrt{3N^2+4h^2/J^2}}{6}\,.
\end{equation}
For the minimum, we require
\begin{equation}
    \frac{\partial^2 E}{\partial K^2}>0 \Rightarrow 6J(2K-N)+4h>0\,.
\end{equation}
Evaluating the second derivative at the stationary points,
\begin{equation}
    \left.\frac{\partial^2 E}{\partial K^2}\right|_{{K}^*_{\pm}}=\pm\frac{2J}{N-1}\sqrt{3N^2+\frac{4h^2}{J^2}}>0\,.
\end{equation}
Therefore, the continuous minimum corresponds to
\begin{equation}
    {K}^*_{+}=\frac{N}{2}-\frac{h}{3J}+\frac{\sqrt{3N^2+4h^2/J^2}}{6}\,.
\end{equation}
Since $K$ must be an integer, the ground-state solution is obtained by comparing the energies of the two neighboring integers, $\lfloor{K}^*_{+}\rfloor$ and $\lceil{K}^*_{+}\rceil\,$.
\qed

\section{Closed-form expressions for integer-valued solutions}
\label{app:integer_k}

In this appendix, we derive the closed-form expression for all $N$ for which the solution to the GMFI problem $K^*$ in Eq.~\eqref{eq:IsingGroundState} is an integer, as summarized in Section~\ref{sec:ExactKStar}. For recap, we consider the scaling factors $h=J=1$. The solution to the GMFI problem then reads
\begin{align}
    \label{eq:Em0m1minkstarAppendix}
    K^* = \frac{3N-2+\sqrt{3N^2+4}}6\,.
\end{align}

The values of $K^*$ which are positive integers have a closed form expression, which we will derive using algebraic number theory. First we note that Eq.~\eqref{eq:Em0m1minkstarAppendix} can only be an integer if $\sqrt{3N^2 + 4}$ is an integer. Therefore we first classify the values $N$ for which $3N^2 + 4$ is a perfect square. This is a standard problem in algebraic number theory.

We consider the number field $\Q(\sqrt{3})$ which has a designated subring the so called {\it ring of integers} $\mathcal{O}$ that contains precisely the arithmetic information we need to answer the question when $3N^2 + 4$ is a perfect square. Indeed, the following properties, which can all be found in \cite{neukirch}, hold:

\begin{itemize}
    \item The ring of integers $\mathcal{O}$ is a free $\Z$-module of rank $2$ which is generated by $1$ and $\sqrt{3}$, i.e. each element $x \in \mathcal{O}$ can be written as $x = a + b\sqrt{3}$ for some $a,b \in \Z$.
    \item The norm map $N: \mathcal{O} \rightarrow \Z$ given by $N(a + b\sqrt{3}) = a^2 - 3b^2$ is multiplicative which means that for two $x,y \in \mathcal{O}$ we have that $N(xy) = N(x)N(y)$.
    \item From the last item it is easy to conclude that an $x \in \mathcal{O}$ is a unit if and only if $N(x) = \pm 1$.
    \item The group of units of $\mathcal{O}$ which we denote by $\mathcal{O}^*$ is a free abelian group of rank $1$ and generated by the fundamental unit $\epsilon = 2+\sqrt{3}$.
\end{itemize}

Let us now suppose that $3N^2+4 = \ell^2$ is a perfect square, then rearranging the equation and setting $a = \ell$ and $b=N$ gives

\begin{equation}\label{eq:pell-equation}
    a^2-3b^2 = 4.
\end{equation}

This equation is an instance of Pell's equation and its solutions are well studied, see again for instance~\cite{neukirch}. In order to solve Eq.~\eqref{eq:pell-equation} we will need to find elements $x \in \mathcal{O}$ with $N(x) = 4$. The following lemma classifies all those $x$.

\begin{lemma}\label{lem:correspondence-pell}
    Let $x \in \mathcal{O}$. We have $N(x) = 4$ if and only if $x = 2\epsilon^n$ for $n \in \Z$.
\end{lemma}

\begin{proof}
    Since $\epsilon$, and more general $\epsilon^n$ for any $n \in \Z$, is a unit, we have that $N(\epsilon^n) = 1$. Furthermore, as the norm is multiplicative we have that $N(2\epsilon^n) = N(2)N(\epsilon^n) = 4$.

    Now we are only left to prove the converse, i.e. that every solution to $N(x)=4$ is of the form $x= 2\epsilon^n$. So let $x = a + b\sqrt{3} \in \mathcal{O}$ with $a,b \in \Z$ be any solution to $N(x)=4$. We will show that $a$ and $b$ are divisible by $2$. It then follows that as well $x/2 \in \mathcal{O}$ and by the multiplicativity of the norm we have
    \[
        N(x) = N(2)N(x/2)
    \]
    from which it follows that $N(x/2) = 1$. But by the above this implies that $x/2$ is a unit and hence of the form $\varepsilon^n$ for some $n \in \Z$ which proves the desired shape for $x$. Therefore, to finish the proof, we now prove that $a$ and $b$ are both even. The easiest way to see this is to reduce
    \begin{equation}\label{eq:normx}
        N(x) = a^2 - 3b^2 = 4
    \end{equation}
    modulo $2$ and find that $a^2 \equiv b^2 \pmod{2}$ and thus $a \equiv b \pmod{2}$, so $a$ and $b$ have the same parity. Now suppose that $a$ and $b$ are both odd. This means that $a$ and $b$ modulo $4$ can each take either the value $1$ or $3$. But none of these four combinations provides a solution to Eq.~\eqref{eq:normx} modulo $4$. Hence, $a$ and $b$ must be both even integers, which finishes the proof.
\end{proof}
Note that if $2\epsilon^n = a + b \sqrt{3}$ for some $n \geq 1$, then $2\epsilon^{-n} = a - b \sqrt{3}$. Since $b$ in our case always corresponds to $N$ which is positive, we will in the subsequent discussion ignore the solutions to $N(x)=4$ coming from negative $n$.

We will use this information to generate a closed form expression for those $K^*$ which are integral and provide a fast way to compute them.
The argument is standard and will be similar to the well-known example of Fiboncacci numbers, as we will start by deriving a second order recurrence for all $a_n$ and $b_n$ that satisfy Eq.~\eqref{eq:pell-equation}. The next pair $(a_{n+1},b_{n+1})$ can be generated from the previous one $(a_n,b_n)$ by
\[
    a_{n+1} + b_{n+1}\sqrt{3} = (a_n + b_n\sqrt{3})(2+\sqrt{3}).
\]
The above lemma implies that once we start with the base case we will exhaust all solutions with this method. Indeed, it follows that
\[
    a_{n+1} = 2 a_n + 3b_n, \quad b_{n+1} = 2b_n + a_n.
\]
We can substitute either equation into the other and obtain the following second order linear recurrences
\[
    a_{n+1} = 4a_n - a_{n-1}, \quad b_{n+1} = 4b_n - b_{n-1},
\]
so both sequences follow the same recurrence with the only difference being the initial values, namely $a_0 = 2, a_1 = 4$ and $b_0 = 0, b_1 = 2$. We set
\[
    A = \begin{pmatrix} 4 & -1 \\ 1 & 0 \end{pmatrix}.
\]
We can then conveniently write
\[
    \begin{pmatrix}
        a_{n+1} \\ a_n
    \end{pmatrix}
     = A^n \begin{pmatrix} 4 \\ 2 \end{pmatrix}, \quad \begin{pmatrix}
        b_{n+1} \\ b_n
    \end{pmatrix}
     = A^n \begin{pmatrix} 2 \\ 0 \end{pmatrix}.
\]

From this we infer an $O(\log n)$ algorithm to compute $a_n$ and $b_n$ efficiently. By diagonalizing $A$ it is also straightforward to derive the following closed-form expressions:
\[
    a_n = \epsilon^n + \epsilon^{-n},\quad b_n = \frac{\epsilon^n - \epsilon^{-n}}{\sqrt{3}}.
\]

We will now answer the question when $K^*$ is an integer. Recall from Eq.~\eqref{eq:Em0m1minkstar} that
\[
    K^* = \frac{3N-2+\sqrt{3N^2+4}}6.
\]
We now insert any solution $(N, \sqrt{3N^2+4}) = (b_n, a_n)$ and find
\[
    K^* = \frac{3b_n - 2 + a_n}{6}.
\]
By the Chinese Remainder Theorem we need to find those $n$ such that $3b_n - 2 + a_n$ is divisible by $2$ and by $3$. However, we see immediately from the recurrence relation that $a_n$ and $b_n$ are always even, so for all $n$ we have
\[
    3b_n - 2 + a_n \equiv 0 \pmod{2}.
\]
Therefore, we only need to check the congruence modulo $3$. We note that $A^6 \equiv I \pmod{3}$ and therefore the $a_n$ and $b_n$ are periodic with period $6$ modulo $3$. Since $3b_n \equiv 0 \pmod{3}$ we only need to find those $n$ such that $a_n \equiv 1 \pmod{3}$. By checking the first $6$ values we find that this congruence is satisfied for any even $n$. Thus, the full sequence of $N_n$ such that $K^*$ is an integer is given by
\[
    N_n^* = b_{2n} = \frac{\epsilon^{2n} + \epsilon^{-2n}}{\sqrt{3}} = \frac{(7+4\sqrt{3})^{n} + (7-4\sqrt{3})^{n}}{\sqrt{3}}.
\]
For computational purposes the number $N_n$ can be computed in $O(\log n)$ operations from the recursion given above. In a similar way we can now derive closed form expressions for those $K^*$ which are integers. We collect the important results in the following Theorem.

\begin{theorem}
    \label{thm:Kinteger}
    The complete classification of pairs $(N,K^*)$ such that $K^*$ is a positive integer is given by the following discrete sequence of pairs $(N_n^*, K_n^*)_{n \geq 0}$ where $N_n^*$ and $K_n^*$ have the following closed form algebraic expressions
    \begin{align}
        N_n^* &= \frac{(7 + 4\sqrt{3})^n - (7 - 4\sqrt{3})^n}{\sqrt{3}}, \\
        K_n^* &= \frac{(1-\sqrt{3})(7-4\sqrt{3})^n + (1+\sqrt{3})(7+4\sqrt{3})^n - 2}{6}.
    \end{align}

    Furthermore, for computational purpose the following formulas hold which allow a computation of $(N_n^*, K_n^*)$ in $O(\log n)$ operations:
    \begin{align}
        N_n^* &= \begin{pmatrix}
            0 & 1
        \end{pmatrix}\begin{pmatrix}
            4 & -1 \\ 1 & 0
        \end{pmatrix}^{2n} \begin{pmatrix}
            2 \\ 0
        \end{pmatrix}, \\
        K_n^* &= \begin{pmatrix}
            0 & 1
        \end{pmatrix} \begin{pmatrix}
            4 & - 1 \\ 1 & 0
        \end{pmatrix}^{2n} \begin{pmatrix}
            \frac{5}{3} \\[0.25em] \frac{1}{3}
        \end{pmatrix} - \frac{1}{3}.
    \end{align}
\end{theorem}

\begin{proof}
    The only claims that are left to prove concern the sequence $K_n^*$. The closed form expression follows immediately by plugging in the expressions of $a_n$ and $b_n$ derived before in the formula for $K_n^*$.

    If we insert the recurrence of $a_n$ and $b_n$ into the sequence
    \[
        c_n = \frac{3b_n - 2 + a_n}{6},
    \]
    we find that $c_n$ satisfies the recurrence $c_{n+1} = 4c_n - c_{n-1} + \frac{2}{3}$ with initial values $c_0 = 0, c_1 = \frac{4}{3}$. We make the change of variables $d_n = c_n + \frac{1}{3}$ to remove the constant term. The initial values evaluate to $d_0 = \frac{1}{3}, d_1 = \frac{5}{3}$. It follows that 
    \[
        d_n = \begin{pmatrix}
            0 & 1
        \end{pmatrix} \begin{pmatrix}
            4 & - 1 \\ 1 & 0
        \end{pmatrix}^n \begin{pmatrix}
            \frac{1}{3} \\[0.25em] \frac{5}{3}
        \end{pmatrix}.
    \]
    Now everything follows by shifting back $c_n = d_n - \frac{1}{3}$ and observing that $K_n^* = c_{2n}.$
\end{proof}

\begin{table}
\centering
\caption{\label{tab:Kinteger}First values of $N_n^*$ and $K_n^*$, for which the solution to the GMFI ground-state Eq.~\eqref{eq:IsingGroundState} is a positive integer. These values are given by the closed-form expressions proven in Theorem~\ref{thm:Kinteger}.}
\begin{ruledtabular}
\begin{tabular}{rrr}
$n$ & $N_n^*$ & $K_n^*$ \\
\hline
0  & 0            & 0 \\
1  & 8            & 6 \\
2  & 112          & 88 \\
3  & 1560         & 1230 \\
4  & 21728        & 17136 \\
5  & 302632       & 238678 \\
6  & 4215120      & 3324360 \\
7  & 58709048     & 46302366 \\
8  & 817711552    & 644908768 \\
9  & 11389252680  & 8982420390 \\
10 & 158631825968 & 125108976696 \\
\end{tabular}
\end{ruledtabular}
\end{table}

\section{Closed-form expressions for the degenerate cases}
\label{app:degeneracy_proof}

Here, we derive the complete sequence of system sizes (Eq.~\eqref{eq:degenerate_system_sizes}) for which the GMFI model exhibits a degenerate ground state. 

\begin{theorem}
    \label{thm:degeneracies}
    The complete classification of degenerate GMFI cases $(N,K^{+})$ that satisfy Eq.~\eqref{eq:DegeneracyEquation} is given by the following discrete sequence of pairs $(N_n^+, K_n^+)_{n \geq 0}$ where $N_n^+$ and $K_n^+$ have the following closed form algebraic expressions
    \begin{align}
        N_n^+ &= \frac{(2+\sqrt{3})^{2n+1} - (2 - \sqrt{3})^{2n+1}}{2\sqrt{3}}, \\
        K_n^+ &= \frac{(1-\sqrt{3})(2-\sqrt{3})^{2n+1} + (1+\sqrt{3})(2+\sqrt{3})^{2n+1} - 10}{12}.
    \end{align}

    Furthermore, for computational purpose the following formulas hold which allow a computation of $(N_n^+, K_n^+)$ in $O(\log n)$ operations:
    \begin{align}
        N_n^+ &= \begin{pmatrix}
            0 & 1
        \end{pmatrix}\begin{pmatrix}
            4 & -1 \\ 1 & 0
        \end{pmatrix}^{2n+1} \begin{pmatrix}
            1 \\ 0
        \end{pmatrix}, \\
        K_n^+ &= \begin{pmatrix}
            0 & 1
        \end{pmatrix} \begin{pmatrix}
            4 & - 1 \\ 1 & 0
        \end{pmatrix}^{2n+1} \begin{pmatrix}
            \frac{5}{6} \\[0.25em] \frac{1}{6}
        \end{pmatrix} - \frac{5}{6}.
    \end{align}
\end{theorem}

\begin{proof}
The proof is completely analogous to the argument given in Section~\ref{sec:ExactKStar} and App.~\ref{app:integer_k}. Note that this time we are asking when $3N^2+1$ is a perfect square. 
By Lemma \ref{lem:correspondence-pell} we know that all the solutions to $N(x) = 4$ are in bijection with $\epsilon^n$, where the bijection is multiplication by $2$. In particular, if $a_n$ and $b_n$ are the sequences from App.~\ref{app:integer_k}, then we can define
\[
    a_n^+ = \frac{a_n}{2}, \quad b_n^+ = \frac{b_n}{2}
\]
and everything can be translated into the situation where we look for degeneracies, i.e. finding $N$ such that $K^+$ is a positive integer. Just note that this time the Chinese Remainder Theorem gives that
\[
    K^+ = \frac{3b_n^+ - 5 + a_n^+}{6}
\]
is an integer if and only if $n$ is odd. The remainder of the proof is routine verification with the techniques presented in App.~\ref{app:integer_k}.
\end{proof}

\begin{table}
\centering
\caption{\label{tab:degeneracies}First values of $N_n^+$ and $K_n^+$, for which the ground state of the GMFI model is degenerate. These values are given by the closed-form expressions proven in Theorem~\ref{thm:degeneracies}.}
\begin{ruledtabular}
\begin{tabular}{rrr}
$n$ & $N_n^+$ & $K_n^+$ \\
\hline
0  & 1            & 0 \\
1  & 15           & 11 \\
2  & 209          & 164 \\
3  & 2911         & 2295 \\
4  & 40545        & 31976 \\
5  & 564719       & 445379 \\
6  & 7865521      & 6203340 \\
7  & 109552575    & 86401391 \\
8  & 1525870529   & 1203416144 \\
9  & 21252634831  & 16761424635 \\
10 & 296011017105 & 233456528756 \\
\end{tabular}
\end{ruledtabular}
\end{table}

Note that the Chebyshev polynomials $T_n(x)$ and $U_n(x)$ are defined by the recurrence relations (see \cite{abramowitz})
\begin{align}
    T_{n+1}(x) &= 2xT_n(x) - T_{n-1}(x),\quad T_1(x) = x,\quad T_0(x) = 1, \\
    U_{n+1}(x) &= 2xU_n(x) - U_{n-1}(x), \quad U_1(x) = 2x, \quad U_0(x) = 1.
\end{align}

We note that for $x=2$ we have the same recurrence relation as for 
the sequences $a_n$ and $b_n$ from App.~\ref{app:integer_k} just with different initial values. As $N_n^*$ and $N_n^+$ are directly derived from these sequences, they are directly related to the values of the Chebyshev polynomials. Indeed, we have for example $N_n^* = 2U_{2n-1}(2)$ and $N_{n}^+ = U_{2n}(2)$.

Thus, the pairs $(N_n^+,K_n^+)$ completely characterize the system sizes and domain-wall positions for which the GMFI model has a degenerate ground state.

\section{Classical energy gap}
\label{appendixclassicalgap}
To compute the classical energy gap, i.e., the difference between the lowest two values of the energy function $E(m_0,m_1)$ in Eq.~\eqref{eq:Em0m1}, we take a closer look at the classical energy landscape around the minimum at $(m_0^*,m_1^+(m_0^*))$ (red downward-pointing triangle in Fig.~\ref{fig:energy_landscape}(b)). Since the energy increases for both $m_0<m_0^*$ and $m_0>m_0^*$ (cyan curve in Fig.~\ref{fig:energy_landscape}(b)), and also along $m_1<m_1^+(m_0^*)$ (gray line in Fig.~\ref{fig:energy_landscape}(b)), there are in general only four grid points around the minimum that qualify for the lowest and second-lowest energy. 

These grid points can be easily enumerated in terms of the transformed coordinates $(K=K(m_0),T=T(m_1))$ (see Eqs.~\eqref{eq:K} and~\eqref{eq:T}) and are given by
\begin{equation}
\begin{aligned}
    S=\{&(\lfloor K^*\rfloor, T^+(\lfloor K^*\rfloor)),\\
    &(\lceil K^*\rceil, T^+(\lceil K^*\rceil)),\\
    &(\lfloor K^*\rfloor, T^+(\lfloor K^*\rfloor)-1),\\
    &(\lceil K^*\rceil, T^+(\lceil K^*\rceil)-1)\}\,,
    \label{eq:fourgridpoints}
\end{aligned}
\end{equation}
where $K^*$ is given by Eq.~\eqref{eq:Em0m1minkstar} and $T^+(K)$ is given by Eq.~\eqref{eq:TplusofK}. Note that in the set $S$, equal grid points for which $\lfloor K^* \rfloor = \lceil K^* \rceil$ (see Section~\ref{sec:ExactKStar} and App.~\ref{app:integer_k}) automatically collapse.
We can thus compute the classical gap by evaluating the energy $E(K,T)$ (see Eq.~\ref{eq:EKT}) for all points in $S$ and taking the absolute difference of the lowest and second-lowest values. Of course, in the special cases $N=15, 209, \ldots$ in which the GMFI model is degenerate (see Table~\ref{tab:degeneracies}), one has to be careful in the numerical evaluation of the energies.

The only exception to the rule given by Eq.~\eqref{eq:fourgridpoints} is the special case $N=3$, in which case the above argument does not apply because the grid point $(\lceil K^*\rceil, T^+(\lceil K^*\rceil)-1)$ does not exist. In this case, the two lowest-energy states are given by $\mid\downarrow\uparrow\uparrow\rangle$ corresponding to $E(K=2,T=3)=-3$ and $\mid\downarrow\downarrow\uparrow\rangle$ corresponding to $E(K=1,T=2)=-1$ with a classical energy gap of $2$. 

\section{Embedding problems onto D-Wave quantum annealers}
\label{app:embeddings}
To solve an Ising problem on a quantum annealer, the problem variables need to be mapped to the topology of the hardware.
Sometimes, when the connectivity of the hardware graph is sufficient to cover all the connections between the problem variables, i.e., the connectivity of the problem describes a subgraph of the hardware topology, the problem can be directly embedded, and it is sufficient to relabel the variables.
However, often the connectivity of the problem is higher than the connectivity of the hardware, i.e., the connectivity of the problem describes a graph which is no longer a subgraph of the hardware topology.
In this case, it is not possible to find a direct embedding of the problem onto the hardware.
To still be able to use the quantum annealer, minor embedding is used which may map a single problem variable (a logical qubit) to a chain of physical qubits.
This chain of physical qubits then has enough connections for the original problem but also has to behave like a single variable, i.e., all the physical qubits in a chain have to have the same value to represent a valid state for the logical problem variable.
In order to achieve this, qubits in a chain are strongly ferromagnetically coupled, favoring the same alignment of the qubits.

\begin{figure}[tb]
    \centering
    \includegraphics[width=\columnwidth]{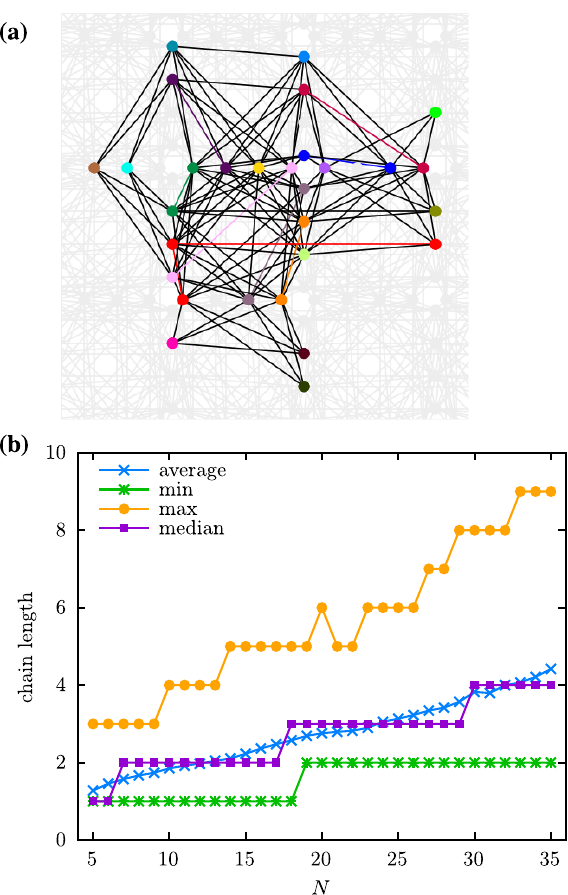}
    \caption{(a) Example embedding of the problem with size $N=10$. The total number of qubits is 29: 10 ancilla qubits are needed to emulate the linear biases $h_i$ using the coupling strengths $J_{i,i+N}$, and 19 qubits are needed to achieve all the couplings between the 10 logical variables. Qubit chains and their connections have the same color, connections of the problem are black, gray qubits and connections show the underlying Zephyr graph where the problem is embedded to. (b) Statistics of the generated embeddings excluding the ancilla qubits $N$, ... $2N-1$ for emulating the linear biases. Average chain length (blue crosses), minimum chain length (green asterisks), maximum chain length (orange circles) and median chain length (purple squares) are shown as a function of problem size $N$.} 
    \label{fig:embedding_statistics}
\end{figure}

\begin{figure*}[t]
    \centering
    \includegraphics[width=\textwidth]{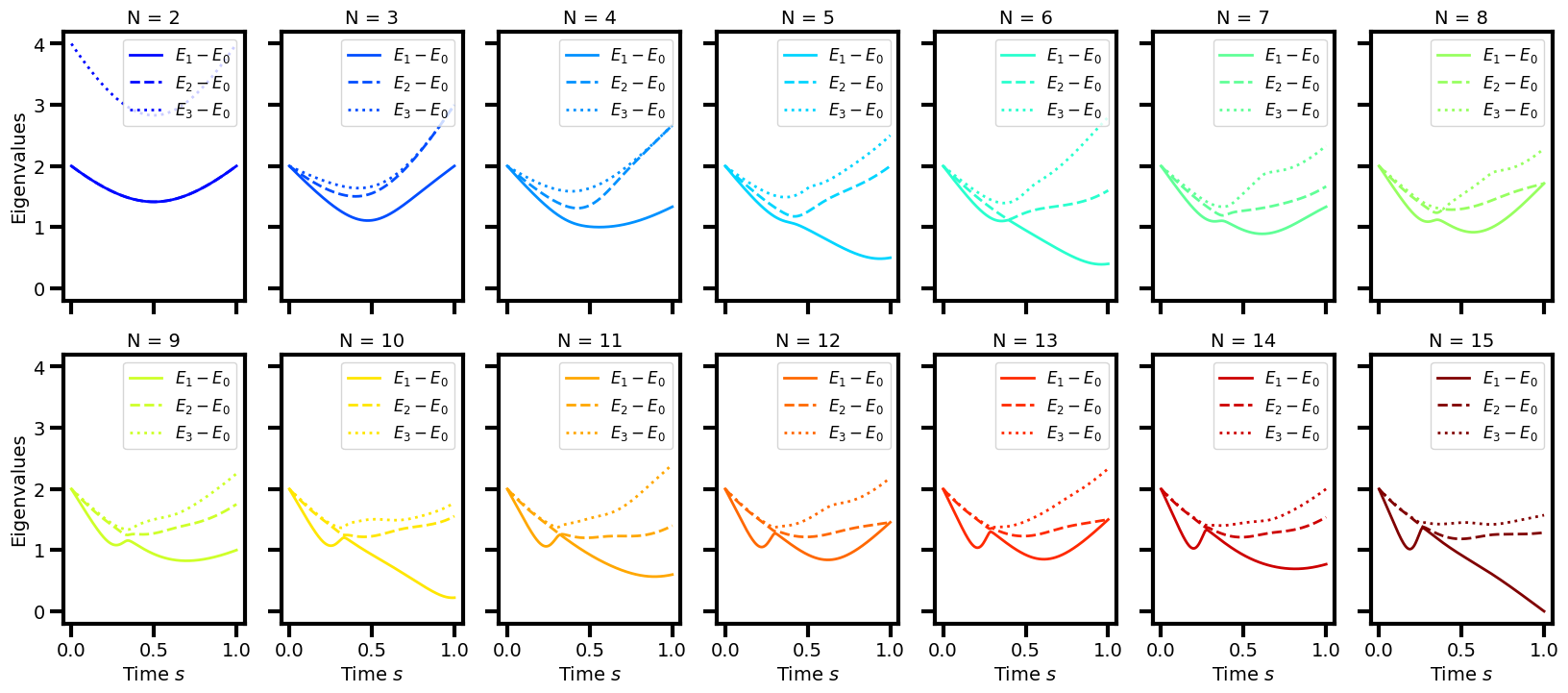}
    \caption{Low-energy spectrum of the GMFI model during quantum annealing with a linear schedule. 
    Shown are the differences $E_i-E_0$ of the lowest $i=1,2,3$ excited-state energies to the ground-state energy $E_0$ as a function of the normalized time $s$ (see legend) for $2\le N \le15$ spins.
    The energies are obtained by diagonalizing the quantum annealing Hamiltonian $H(s)$ in Eq.~\eqref{eq:QAHamiltonian} at 300 different linearly-spaced times $s\in[0,1]$. The data for the spectral gaps $E_1-E_0$ (solid lines) agree with the data shown in Fig.~\ref{fig:spectral_gaps} of the main text for the corresponding $N$.
    The rightmost figures for $N=8$ ($N=15$) correspond to the maximal (minimal / degenerate) cases of the second gap (cf.~Fig.~\ref{fig:theoretical_quantum_annealing}c).
    }
    \label{fig:qa_low_energy_spectrum}
\end{figure*}

In theory, this ferromagnetic coupling (also called chain strength) can be made arbitrarily strong so that the ground state of the embedded problem corresponds to the ground state of the original problem.
In practice, the chain strength cannot be made arbitrarily large as the $J$ and $h$ values are limited by range and precision.
So if the chain strength is too large, the problem parameters $J_{ij}$ and $h_i$ need to be scaled down to fit into the range limits, leading to a loss in precision and thus noisier results which do not necessarily correspond to low-energy states of the problem anymore. 
If the chain strength is not strong enough, the ground state of the embedded problem may correspond to an invalid configuration where physical qubits representing the same logical variable attain different values which cannot directly be mapped to a valid state of the original problem.
In such a case where qubits of a single chain take different values (also called chain break), post-processing methods can fix these to yield a valid state of the original problem, however such a state does not need to be a low-energy state or the ground state of the problem.

Our GMFI problems have a high connectivity so that direct embedding onto the Zephyr topology is not possible and minor embedding is needed.
As described in the main text, we do not want the problem to be scaled-down, so we accept the chain breaks and discard these samples as we do not want the post-processing to alter the success rates either.
We show the average, minimum, maximum and median chain lengths of the generated embeddings as a function of problem size $N$ in Fig.~\ref{fig:embedding_statistics}.

\section{Low-energy quantum annealing spectrum}
\label{appendixlowenergy}
In Fig.~\ref{fig:qa_low_energy_spectrum}, we show the lowest eigenenergies of the TFIM quantum annealing Hamiltonian $H(s)$ given by Eq.~\eqref{eq:QAHamiltonian}.

\end{document}